\documentclass[letterpaper, 11pt]{article}
\pdfoutput=1

\usepackage{jheppub}

\usepackage{graphicx}
\usepackage{epstopdf}
\usepackage{amsmath, amssymb}

\usepackage{bm}
\usepackage{caption}
\usepackage{subcaption}

\usepackage{multirow} 
\usepackage{longtable} 

\usepackage{romannum}

\usepackage[makeroom]{cancel} 

\usepackage{dsfont}

\usepackage{comment}

\newcommand{\be}{\begin{eqnarray}}
\newcommand{\ee}{\end{eqnarray}}
\newcommand{\nn}{\nonumber}
\newcommand{\bn}{\begin{enumerate}}
\newcommand{\en}{\end{enumerate}}

\def\CI{{\cal I}}

\def\CN{{\cal N}}

\def\CT{{\cal T}}

\def\CZ{{\cal Z}}

\def\a{\alpha}
\def\b{\beta}
\def\g{\gamma}

\def\r{\rho}

\def\half{\frac{1}{2}}

\def\vev#1{\langle #1 \rangle}

\def\vec#1{\bm{#1}}

\def\nn#1{\mathcal{N}=#1}

\newcommand{\bea}{\begin{eqnarray}}
\newcommand{\eea}{\end{eqnarray}}

\def\CI{{\cal I}}

\def\CN{{\cal N}}

\def\CT{{\cal T}}

\def\CZ{{\cal Z}}

\def\a{\alpha}
\def\b{\beta}
\def\g{\gamma}

\def\r{\rho}

\def\half{\frac{1}{2}}

\def\vev#1{\langle #1 \rangle}

\allowdisplaybreaks

\title{On dimensional reduction of 4d N=1 Lagrangians for Argyres-Douglas theories}

\author{Prarit Agarwal}
\affiliation{Department of Physics and Astronomy \& Center for Theoretical Physics\\ Seoul National University, Seoul 151-747, Korea}

\abstract
{

Recently, it was found that certain 4d $\nn{1}$ Lagrangians experience supersymmetry enhancement at their IR fixed point, thereby giving a Lagrangian description for a plethora of Argyres-Douglas theories. A generic feature of these Lagrangians is that a number of gauge invariant operators decouple (as free fields) along the RG-flow. These decoupled operators can be naturally taken into account from the beginning itself by introducing additional gauge singlets (sometimes called ``flipping fields'') that couple to the decoupled operators via appropriate superpotential terms. It has also been checked that upon dimensionally reducing to 3d, the $(A_1,A_{2n-1})$ type Lagrangians only produce the expected behavior when flipping fields are included in the Lagrangian. In this paper we further investigate the role of flipping fields and find an example where the expected necessity of including the flipping fields in the dimensionally reduced Lagrangians seems to get violated.  In the process we find two new dual Lagrangians for the so called 3d $T[SU(2)]$ theory.    

}

\preprint{SNUTP18-004}

\begin{document}
\maketitle

\section{Introduction}

It is a generic fact that quantum field theories (QFTs) become scale invariant at low energies. In all known cases of unitary QFTs, scale invariance is also accompanied by invariance under the  ``special conformal transformations''. It therefore follows that QFTs flow to conformal field theories (CFTs) in the infrared (IR). It is therefore no overstatement that understanding CFTs in general is an imperative cause being pursued by physicists all over the world. 

Developments in string and M-theory have also uncovered the existence of the so called ``non-Lagrangian theories''. These are CFTs which can be probed through their string theoretic construction, however, their quantum excitations are very strongly coupled, such that they have no known Lagrangian description.  Amongst the earliest (and perhaps the simplest) examples of such non-Lagrangian theories is the Argyres-Douglas (AD) theory \cite{Argyres:1995jj}. This is also referred to as the $H_0$ theory in literature. It is a 4d CFT with $\nn{2}$ supersymmetry and can be thought of as the theory describing the electromagnetic interaction between a $U(1)$ magnetic monopole and a dyon, along with their superpartners. Argyres and Douglas discovered it by studying a very specific point on the Coulomb branch of the 4d $\nn{2}$ $SU(3)$ super-Yang-Mills theory. This description in terms of a critical point on the Coulomb branch of a Lagrangian theory makes it possible to study the Coulomb phase of the AD-theory, however, its conformal phase is not as easily accessible through such a description.   

By now, we know an infinite set of 4d $\nn{2}$ supeconformal field theories (SCFTs) that can be considered close cousins of the $H_0$ theory in that their Coulomb phase consists of a system of interacting particles with mutually non-local electromagnetic charges \cite{Argyres:1995xn,Eguchi:1996ds,Eguchi:1996vu,Gaiotto:2009hg,Bonelli:2011aa,Gaiotto:2012sf,Xie:2012hs,Cecotti:2010fi,Xie:2013jc,Wang:2016yha,Wang:2018gvb}. These are referred to as generalized AD theories. Owing to the work of \cite{Shapere:2008zf}, a prescription to compute their central charges is also available. Based on the general connection betweeen 4d $\nn{2}$ SCFTs and 2d chiral algebras \cite{Beem:2013sza}, the authors of \cite{Liendo:2015ofa} were able to show that the central charge $c$ of the $H_0$ theory saturates a lower bound and therefore in this sense the AD theory can be thought of as the simplest of all 4d $\nn{2}$ SCFTs.

A quantity that is very useful to characterize the spectra of an SCFT is its superconformal index. The lack of Lagrangian description for generalized AD theories implies that their superconformal index is not readily computable. However, for many classes of the generalised AD theories, the insights of \cite{Buican:2015ina, Cordova:2015nma, Song:2015wta, Buican:2015tda,Buican:2017uka, Song:2017oew} have made it possible to compute the superconformal index in the so called Schur and Macdonald limits. 

The full $\nn{2}$ superconformal index of the $H_0$ theory was first computed in \cite{Maruyoshi:2016tqk}. This was made possible through the discovery of a 4d $\nn{1}$ Lagrangian theory which undergoes SUSY enhancement at the end of its RG flow with its fixed point being the $H_0$ theory. 4d $\nn{1}$ Lagrangians for many other generalized AD theory were obtained in \cite{Maruyoshi:2016aim,Agarwal:2016pjo,Agarwal:2017roi,Benvenuti:2017bpg}. For all those generalized AD theories whose 4d $\nn{1}$ Lagrangians are known, the computation of their full superconformal index is therefore a straightforward exercise.  Besides being inherently interesting owing to being rare examples of SUSY enhancing 4d RG flows \footnote{We passingly note that 4d $\nn{1}$ Lagrangians that flow to the $E_6$ and $E_7$ Minahan-Nemeschansky theory \cite{Minahan:1996fg,Minahan:1996cj} were obtained in \cite{Gadde:2015xta} and \cite{Agarwal:2018ejn} respectively. However, their construction is different from that of the Lagrangians for AD theories. }, these $\nn{1}$ Lagrangians have also been successfully used to probe many other properties of the generalized AD theories \cite{Cordova:2017mhb,Cornagliotto:2017snu,Gukov:2017zao,Dedushenko:2018bpp,Fluder:2017oxm}.  

One can also consider the 3d $\nn{4}$ SCFTs obtained by reducing AD theories on a circle. These are best described in terms of their mirror duals \cite{Intriligator:1996ex} and were obtained in \cite{Nanopoulos:2010bv,2008arXiv0806.1050B,Benini:2010uu,Xie:2012hs}. It is natural to expect that the dimensional reduction of 4d $\nn{1}$ Lagrangians of generalized AD theories should give 3d Lagrangians whose IR fixed point is described by the corresponding 3d mirrors of AD theories. The authors of \cite{Benvenuti:2017lle,Benvenuti:2017kud} considered exactly this question. They argued that certain terms in the 4d $\nn{1}$ Lagrangians dynamically dropout due to chiral ring instability and hence such terms should be removed from the superpotential. Additionally, one can add certain ``flipping fields'' to the Lagrangians. The purpose of these flipping fields is to take into account the decoupling of certain gauge invariant operators by removing them from the chiral ring.  
It was shown in \cite{Benvenuti:2017lle,Benvenuti:2017kud} that the dimensionally reduced $(A_1, A_{2n+1})$ type Lagrangians only flows to the desired fixed point when the flipping fields are included in the dimensionally reduced Lagrangian. Thereby bringing to light the importance of including ''flipping fields'' whenever the RG flow involves decoupled operators. 

However, the addition of ''flipping fields'' is a rule-of-thumb that, while expected to hold in generic cases, might at times fail. In this paper we will demonstrate one such example where the addition of a flipping field to the dimensionally reduced Lagrangian actually spoils the expected match. The example we will present corresponds to the Lagrangian description for the so called $(A_1,D_3)$ AD theory.

The 3d mirror of the dimensionally reduced  $(A_1,D_3)$ theory is given by a 3d $\nn{4}$ $U(1)$ gauge theory coupled to 2 hypers carrying $U(1)$ charge 1. This is popularly called the $T[SU(2)]$ theory. We will show that the dimensionally reduced $(A_1,D_3)$ Lagrangian straightforwardly flows to the $T[SU(2)]$ fixed point without the need to add a flipping field. On the other-hand, addition of a flipping field deforms the Lagragian in such a way that it flows to a completely different fixed point with only 3d $\nn{2}$ supersymmetry.

At this point we will like to point out that the $(A_1,D_3)$ AD theory is identical to the $(A_1,A_3)$ theory. However, depending upon whether one classifies it as part of the $(A_1,A_{2n+1})$ type of AD theories or the $(A_1,D_{2n+1})$ type of AD theories, one can write down two distinct $\nn{1}$ Lagrangians for them \cite{Maruyoshi:2016aim,Agarwal:2016pjo}. In order to distinguish between the two Lagrangians we will call one of them as the $(A_1, D_3)$ Lagrangian and the other as the $(A_1, A_3)$ Lagrangian, respectively. These two Lagrangians are therefore dual to each other in the sense of \cite{Seiberg:1994pq}. 
3d reduction of the $(A_1,A_3)$ Lagrangian was already studied in detail in \cite{Benvenuti:2017lle}. More details about these Lagrangians will be provided in the relevant sections of the paper.  


The plan of the paper is as follows: In section \ref{sec:nildef}, we review the construction of 4d $\nn{1}$ Lagrangians for generalized AD theories. Section \ref{sec:TSU2} is devoted to reminding the reader about some basic facts concerning the $T[SU(2)]$ theory. In section \ref{sec:A1A3}, we summarize   dimensional reduction of the so called $(A_1,A_3)$ Lagrangian, as was first done in \cite{Benvenuti:2017lle}. We also compute the 3d superconformal index of the dimensionally reduced $(A_1,A_3)$ Lagrangian and check that it matches with that of the $T[SU(2)]$ theory. The matching of the superconformal index is the only ingredient in this section that was not considered in \cite{Benvenuti:2017lle}. In section \ref{sec:A1D3}, we consider the dimensional reduction of the $(A_1,D_3)$ Lagrangian and compare it to the $T[SU(2)]$ theory. In section \ref{sec:Mirror}, we consider an alternative derivation of the $T[SU(2)]$ theory as a mirror of the $(A_1,D_3)$ Lagrangian, thereby providing further confirmation of our claim. The Lagrangians thereby described in sections \ref{sec:A1D3} and \ref{sec:Mirror} therefore give us two new duals of the $T[SU(2)]$ theory.

As might be obvious from the plan of the paper mentioned above, sections \ref{sec:nildef}-\ref{sec:A1A3} are general summaries of known facts and can be safely skipped by experts.

\section{Review of 4d $\nn{1}$ Lagrangians for generalized AD theories}
\label{sec:nildef}
The construction of 4d $\nn{1}$ Lagrangians for generalized AD theories is based on a specific kind of $\nn{1}$ preserving deformation of general 4d $\nn{2}$ SCFTs. These were first introduced in \cite{Gadde:2013fma} and further studied in detail in \cite{Agarwal:2013uga,Agarwal:2014rua,Agarwal:2015vla,Nardoni:2016ffl,Fazzi:2016eec,Apruzzi:2018xkw}\footnote{Also see \cite{Benini:2009mz,Bah:2012dg,Bah:2013aha,Xie:2013gma} for the construction of new 4d $\nn{1}$ SCFTs obtained from $\nn{1}$ preserving deformations of the so called class-S theories \cite{Gaiotto:2009we}.} .  The idea is as follows: we start with any given 4d $\nn{2}$ SCFT $\mathcal{T}_{UV}$ having a non-Abelian flavor symmetry $F$. Invariance of $\CT_{UV}$ under $F$  as well as $\nn{2}$ superconformal algebra implies that its spectrum contains a superconformal multiplet of conserved $F$-currents whose lowest component is a scalar, usually called ``the moment map operator'' and denoted by $\mu$. We now deform this theory by introducing a gauge singlet chiral superfield $M$ transforming in the adjoint representation of $F$, and couple it to $\CT_{UV}$ via a superpotential term given by 
\be
\delta W = {\rm Tr} M \mu \ .
\ee 
Furthermore, we give a nilpotent vev to $M$ : $\vev{M}=\rho(\sigma^+)$, where $\rho$ specifies the choice of an $SU(2)_{\rho} \hookrightarrow F$.  This explicitly breaks the $\nn{2}$ supersymmetry enjoyed by $\CT_{UV}$ and generically triggers an $\nn{1}$ preserving RG-flow. The flavor symmetry group of the theory also gets broken to the commutant of $SU(2)_{\rho} \hookrightarrow F$. 

The deformed superpotential is now given by 
\be
\delta W = {\rm Tr}\rho(\sigma^+) \mu_{j=1, m=-1} \ + \ \sum_{j,k} M_{j,-j,k} \mu_{j,j,k} \ ,
\label{eq:defSup}
\ee 
where $M_{j,m,k}$ is the component of $M$ that transforms in the $(2j+1)$-dimensional irrep. of $SU(2)_\rho$ with a spin $m$, while $k$ denotes its quantum numbers with respect to the remnant flavor symmetry. The notation $\mu_{j,m,k}$ can also be understood in a similar way. The above deformation will also break the $SU(2)_R \times U(1)_r$ R-symmetry of $\CT_{UV}$ down to $U(1)_{I_3} \times U(1)_r$, where $U(1)_{I_3}$ is Cartan subgroup of $SU(2)_R$. The 4d $\nn{1}$ R-symmetry $U(1)_R$ is then given by a linear combination of $U(1)_{I_3}$ and $U(1)_r$. This linear combination is a priori unfixed and is determined by using the principle of ``a-maximization'' \cite{Intriligator:2003jj} and its modification \cite{Kutasov:2003iy}. Once the appropriate linear combination has been determined, the resulting $U(1)_R$-charge can be used to compute the central charges of the IR-fixed point using the relations \cite{Anselmi:1997am}
\be
a = \frac{3}{32} (3 {\rm Tr} R^3 - {\rm Tr} R) \ , \\
c= \frac{1}{32} (9 {\rm Tr} R^3 - 5 {\rm Tr} R)
\ee 


Let us now apply the above deformation  to the 4d $\nn{2}$ Lagrangian SCFT consisting of an $SU(N)$ gauge theory coupled to $N_f = 2N$ fundamental flavors. Let us also choose $\rho:SU(2)_{\rho} \hookrightarrow SU(2N)$ to be given by the principle embedding \footnote{The princinple embedding  $SU(2)_{\rho} \hookrightarrow SU(2N)$ sends the $2N$ dimensional irrep. of $SU(2N)$ to the $2N$ dimensional irrep. of $SU(2)_{\rho}$.}. It turns out that upon doing so, the central charges of the resulting IR fixed point match with those of the so called  $(A_1,A_{2N-1})$ AD theory. It is natural to wonder if the IR fixed point is indeed given by the $(A_1,A_{2N-1})$ AD theory. In \cite{Maruyoshi:2016aim}, Maruyoshi and Song provided convincing evidence to show that this is indeed the case. Thereby they were able to obtain an $\nn{1}$ Lagrangian for the $(A_1,A_{2N-1})$ AD theory. One can also consider a similar deformation of the 4d $\nn{2}$ SCFT consisting of the $USp(2N)$ gauge group coupled to $N_f = 4N+4$ fundamental half-hypers and choose $SU(2)_\rho$  to be the principle nilpotent embedding of its $SO(4N+4)$ flavor symmetry. This gives us an $\nn{1}$ Lagrangian for the $(A_1, A_{2N})$ theories. A more detailed study of all possible nilpotent deformations of the above mentioned 4d $\nn{2}$ SCFTs was carried out in \cite{Agarwal:2016pjo}, thereby leading to the discovery of $\nn{1}$ Lagrangians for $(A_1,D_N)$ theories. More recently, \cite{Giacomelli:2018ziv} found a set of necessary conditions for SUSY enhancement of $\nn{1}$ theories obtained through above mentioned deformations of $\nn{2}$ SCFTs. Meanwhile, the authors of \cite{Carta:2018qke} have shown how to use T-branes to describe the above deformations when applied to rank-1 SCFTs, thereby taking the first steps towards their string theory uplift.

For the purpose of this paper, we need to apply the above deformation to the 4d $\nn{2}$ SCFT consisting of the $SU(2)$ gauge theory coupled to 8 fundamental half-hypers. This theory has an $SO(8)$ flavor symmetry. To obtain the $(A_1,A_3)$ Lagrangian we deform the above theory using $SU(4)\subset SO(8)$ that rotates the 4 hypers formed from the 8 half-hypers. Choosing $SU(2)_\rho$ to be the principle embedding of $SU(4)$ then gives us a 4d $\nn{1}$ Lagrangian that flows to the $(A_1,A_3)$ theory. Along the way some of the gauge singlets introduced through our deformation decouple from the interacting theory. The chiral operator ${\rm Tr}\phi^2$, where $\phi$ is the scalar in the 4d $\nn{2}$ $SU(2)$ vector multiplet also decouples as a free field. 

Similarly if one focuses on the full $SO(8)$ flavor symmetry of the UV $\nn{2}$ SCFT mentioned in the previous paragraph, and chooses $\rho:SU(2)_\rho \hookrightarrow SO(8)$ to be given by 
\be
SO(8) &\rightarrow& SU(2)_{\rho} \nonumber \\
\vec{8} &\rightarrow& \vec{5} \oplus \vec{1} \oplus \vec{1}\oplus \vec{1} \ ,
\ee 
then the 4d $\nn{1}$ Lagrangian so obtained flows to the $(A_1,D_3)$ theory. Once again, some of the gauge singlets introduced through our deformation as well as the gauge invariant operator ${\rm Tr}\phi^2$  end up decoupling from the interacting theory. Let us reiterate that while the $(A_1,A_3)$ AD theory and the $(A_1,D_3)$ AD theory are identical, their $\nn{1}$ Lagrangians as described here end up being distinct.



\section{$T[SU(2)]$ theory}
\label{sec:TSU2}

The 3d reduction of the $(A_1,A_3) \equiv (A_1,D_3)$ AD theory is given by the $T[SU(2)]$ theory.  This is an $\mathcal{N}=4$ $U(1)$ gauge theory coupled to 2 hypermutliplets, both having charge 1 with respect to the $U(1)$ gauge transformations. There is a topological global symmetry $U(1)_T$ that arises from the shift symmetry of the dual photon: $\phi \rightarrow \phi + \text{constant}$. It was argued in \cite{Intriligator:1996ex,Gaiotto:2008ak} that this $U(1)_T$ gets enhanced to $SU(2)_T$. Similarly, the Higgs branch of this theory also has an $SU(2)_b$ global symmetry with the two hypermultiplets of the $T[SU(2)]$ together forming a doublet of $SU(2)_b$. The R-symmetry is given by $SO(4)_R$ but in the $\mathcal{N}=2$ language, only the $U(1)_R \times U(1)_q \subset SO(4)_R$ is manifest. We summarize the matter content of this theory in \eqref{tab:TSU2}.
\be
	\centering
	\begin{tabular}{|c|c|c|c|c|}
		\hline
		fields & $U(1)_{\text{gauge}}$& $U(1)_b \subset SU(2)_{b}$ & $U(1)_q$ & $U(1)_R$\\
		\hline \hline
		$P_1$ & $+1$ & $+1$ &$+1$ & $+\frac{1}{2}$ \\ \hline
		$\widetilde{P}_1$ & $-1$ & $-1$ &$+1$ & $+\frac{1}{2}$ \\ \hline
		$P_2$ & $+1$ & $-1$ &$+1$ & $+\frac{1}{2}$ \\ \hline
		$\widetilde{P}_2$ & $-1$ & $+1$ &$+1$ & $+\frac{1}{2}$ \\ \hline
		$\phi$ & 0 & $0$ & $-2$ & 1 \\ \hline
	\end{tabular}
	\label{tab:TSU2}
\ee 
The superpotential is given by 
\be
W= \phi (P_1 \widetilde{P}_1 + P_2 \widetilde{P}_2) \ .
\ee 
The 3 components of $SU(2)_b$ moment map are given by  $\{P_1 \widetilde{P}_2, \ P_1 \widetilde{P}_1 - P_2 \widetilde{P}_2, \ \widetilde{P}_1 P_2\}$. Similarly the moment map of $SU(2)_T$ consists of $\{ \mathfrak{M}^{+}, \ \phi , \ \mathfrak{M}^{-} \}$. Where $\mathfrak{M}^{\pm}$ are monopole operators with magnetic charges $\pm 1$ respectively. Together, these operators form the chiral ring of $T[SU(2)]$. We list their charges with respect to the various global symmetries in  \eqref{tab:TSU2chiralOp}.  
\be
\centering
\begin{tabular}{|c|c|c|c|c|}
	\hline
	Chiral Op. & $U(1)_{T} \subset SU(2)_T$& $U(1)_b \subset SU(2)_{b}$ & $U(1)_q$ & $U(1)_R$\\
	\hline \hline
	$P_1 \widetilde{P}_2$ & $0$ & $+2$ &$+2$ & $+1$ \\ \hline
	$\ P_1 \widetilde{P}_1 - P_2 \widetilde{P}_2$ & $0$ & $0$ &$+2$ & $+1$ \\ \hline
	$\widetilde{P}_1 P_2$ & $0$ & $-2$ &$+2$ & $+1$ \\ \hline \hline
	$\mathfrak{M}^+$ & $1$ &$0$& $-2$ & 1 \\ \hline
	$\phi$ & 0 & $0$ & $-2$ & 1 \\ \hline
	$\mathfrak{M}^-$ & $-1$ &$0$& $-2$ & 1 \\ \hline
\end{tabular}
\label{tab:TSU2chiralOp}
\ee 
It is also quite straight forward to compute the superconformal index and the $S^3$ partition function of the $T[SU(2)]$ theory. We will use these to establish the duality between the $T[SU(2)]$ theory and the 3d Lagrangian obtained from dimensional reduction of the $(A_1,D_3)$ Lagrangian. 

{\noindent\textbf{The 3d Superconformal Index:}} In the absence of any Chern-Simons term the 3d superconformal index (SCI) of a Lagrangian SCFT with a gauge group $G$ can be written as \cite{Kim:2009wb,Bhattacharya:2008zy,Imamura:2011su} \footnote{In Chern-Simons theories, there is an additional term which captures the contribution from the classical action of the monopole + holonomy configuration on $S^2 \times S^1$ . }
\be
I = \displaystyle \sum_{\vec{m}} \oint \prod_{j} \frac{d z_j}{2 \pi i z_j} \frac{1}{|\mathcal{W}(\vec{m})|}Z_{\text{vec}} (z, x, \vec{m}) \prod_{\Phi} Z_{\Phi} (z, x, t, \vec{m}) \ ,
\ee 
where $\vec{m}$ runs over the allowed magnetic charges (modulo Weyl transformations) for the given gauge group \footnote{ Basically, $\vec{m} \in \Gamma^*_{\hat{G}}/\mathcal{W}_{\hat{G}}$, where $\Gamma^*_{\hat{G}}$ is the weight lattice of the dual gauge group $\hat{G}$ and $\mathcal{W}_{\hat{G}}$ is its Weyl group.}. 
$|\mathcal{W}(\vec{m})|$ is the order of the Weyl group of the subgroup of $G$ that is left unbroken by the magnetic fluxes $\vec{m}$. 
 $Z_{\text{vec}} (z, x, m)$ is the contribution from the vector multipets and is defined by
\be
Z_{\text{vec}} (z = e^{i a}, x, \vec{m}) = \prod_{\alpha \in \Delta} x^{-\frac{|\alpha(\vec{m})|}{2}} (1-e^{i \alpha(a)} x^{|\alpha(\vec{m})|}) \ ,
\ee 
Here $\Delta$ is the set of non-zero roots of $G$. Similarly,
\be
 Z_{\Phi} (z, x, t, \vec{m}) = \prod_{\rho \in R_{\Phi}} \Big( x^{(1-r_{\Phi})} e^{-i \rho (a)} \prod_{k}t_k^{-f_k(\Phi)} \Big)^{\frac{|\rho(\vec{m})|}{2}} \frac{(e^{-i \rho (a)} \prod_{k}t_k^{-f_k(\Phi)} x^{|\rho(\vec{m})|+2-r_{\Phi}}; x^2)_{\infty}}{(e^{i \rho (a)} \prod_{k}t_k^{f_k(\Phi)} x^{|\rho(\vec{m})|+r_{\Phi}}; x^2)_{\infty}} \ , \nonumber \\
\ee 
where $R_{\Phi}$ is the representation of chiral field $\Phi$ with respect to the gauge group. 
In the above formula, we have used the standard notation for the q-Pochhammer symbol 
\be
(a;q)_n = \prod_{k=0}^{n-1}(1-a q^k) \ .
\ee 

\noindent\textbf{SCI of $T[SU(2)]$:} For $T[SU(2)]$, the superconformal index takes the following form
\be
I_{T[SU(2)]} = \sum_{m \in \mathbb{Z}}\oint \frac{d z}{2 \pi i z} t^m Z_{\phi} (x,b,v,z) Z_{P_1} (x,b,v,z)  Z_{\widetilde{P}_1} (x,b,v,z) Z_{P_2} (x,b,v,z)  Z_{\widetilde{P}_2} (x,b,v,z) \ , \nonumber \\
\label{eq:TSU2Ind}
\ee 
where  $v, b$ and $t$ are the fugacities for $U(1)_q, SU(2)_b$ and $U(1)_T$ respectively, and
\be
Z_\phi(x,b,v,z) &=& \prod_{k=0}^{\infty} \frac{1- v^2 x^{2k+1}}{1- v^{-2} x^{2k+1}} \ , \\
Z_{P_1}(x,b,v,z) &=& (x^{\frac{1}{2}} z^{-1} b^{-1} v^{-1})^{\frac{|m|}{2}} \prod_{k=0}^{\infty} \frac{1- z^{-1} b^{-1} v^{-1} x^{2k+|m|+\frac{3}{2}}}{1-z b v x^{2k+|m|+\frac{1}{2}}} \ , \\
Z_{\widetilde{P}_1}(x,b,v,z) &=&Z_{P_1}(x,b^{-1},v,z^{-1}) \ , \\
Z_{P_2}(x,b,v,z) &=&Z_{P_1}(x,b^{-1},v,z) \ , \\
Z_{\widetilde{P}_2}(x,b,v,z) &=&Z_{P_1}(x,b,v,z^{-1}) \ . \\
\ee

\noindent Evaluating \eqref{eq:TSU2Ind} explicitly, we find 
\be
I_{T[SU(2)]} &=& 1+x \left(v^2 \chi _3^b+\frac{\chi _3^t}{v^2}\right)+x^2 \left(-1-\chi _3^b-\chi _3^t+v^4 \chi _5^b+\frac{\chi _5^t}{v^4}\right)+ \nonumber \\
&~~~&x^3 \left(\frac{1}{v^2}+v^2-v^2 \chi _5^b-\frac{\chi _5^t}{v^2}+v^6
\chi _7^b+\frac{\chi _7^t}{v^6}\right)+ \nonumber \\
&~~~&x^4 \left(-2+\chi _3^b.\chi _3^t+v^4 \chi _3^b+\frac{\chi _3^t}{v^4}-v^4 \chi _7^b-\frac{\chi _7^t}{v^4}+v^8 \chi _9^b+\frac{\chi _9^t}{v^8}\right)+ \nonumber \\
&~~~&x^5
\Big(-\frac{\chi _3^b.\chi _3^t}{v^2}-v^2 \chi _3^b.\chi _3^t-2 v^2 \chi _3^b-\frac{2 \chi _3^t}{v^2}-v^2 \chi _5^b+v^6 \chi _5^b+ \nonumber\\ 
&~~~&\hspace{3cm}\frac{\chi _5^t}{v^6}-\frac{\chi _5^t}{v^2}-v^6 \chi_9^b-\frac{\chi _9^t}{v^6}+v^{10} \chi _{11}^b+\frac{\chi _{11}^t}{v^{10}}\Big) + \mathcal{O}(x^6) \ . 
\label{eq:TSU2Index}
\ee 
Here $\chi_n^b$ and $\chi_n^t$ represent the characters of the $n$ dimensional irrep. of $SU(2)_b$ and $SU(2)_T$, where $SU(2)_T$ emerges due to enhancement of $U(1)_T$. We note that the index above is invariant under the $\mathbb{Z}_2$ transformation given by 
\be
\mathbb{Z}_2 : v \rightarrow v^{-1}, \ b \leftrightarrow \sqrt{t} \ .
\ee
This is because the $\mathbb{Z}_2$ transformation described above corresponds to the mirror symmetry that exchanges the Higgs and the Coulomb branch of $T[SU(2)]$. The invariance of the index then follows from the fact that $T[SU(2)]$ is self-mirror.

{\noindent\textbf{The $S^3$ partition function:}} Given a QFT with gauge group $G$ and chiral multiplets $\Phi$ transforming in the representation $R_{\Phi}$ of $G$, its $S^3$ partition function of can be written as \cite{Kapustin:2009kz,Jafferis:2010un,Hama:2010av,Hama:2011ea,Imamura:2011wg}
\be
\mathcal{Z}_{S^3} = \frac{1}{|W|} \int_{-\infty}^{\infty} \prod_{i} dz_i \prod_{\alpha \in \Delta^+} 4 \sinh^2 \pi \alpha(\vec{z}) \prod_{\Phi} \prod_{\rho \in R_\Phi} e^{l(1-r_{\Phi} + i \rho(z))} \ ,
\ee 
where, for the time being, we have switched off the background real masses associated to the various flavor symmetries acting on the chiral fields. $|W|$ is the order of the Weyl group of $G$ and $\Delta^+$ denotes the set of positive roots of $G$. The function $l(z)$ is such that $l'(z) = - \pi z \cot \pi z$. This can be integrated with the boundary condition $l(0)=0$, to give 
\be
l(z)= -z \log (1- e^{2 \pi i z}) + \frac{i}{2} \Big[\pi z^2 + \frac{1}{\pi} {\rm Li}_2 (e^{2 \pi i z}) \Big] - \frac{ \pi i}{12} \ .
\ee 

\noindent For the case of $T[SU(2)]$, the $S^3$ partition function then becomes
\be
\mathcal{Z}_{S^3 , T[SU(2)]} = \int_{-\infty}^{\infty} dz e^{2l(\half + i z)+2l(\half - i z)} = \frac{1}{2\pi}\ .
\ee 
 
\section{The $(A_1, A_3)$ Lagrangian}
\label{sec:A1A3}
In this section we will describe the $(A_1,A_3)$ Lagrangian.
Its dimensional reduction to 3d was first considered in \cite{Benvenuti:2017lle}.
 Let us quickly review the 4d Lagrangian, followed by its 3d reduction. The Lagrangian consists of an $SU(2)_{\text{color}}$ gauge theory with 2 chiral multiplets, $q_m, \ m \in \{1,2\}$, each transforming in the doublet representation of $SU(2)_{\text{color}}$. In addition to this there is another chiral superfield $\phi$ that transforms in the adjoint irrep of $SU(2)_{\text{color}}$ and gauge singlets $M_3, \beta$ which are coupled to the rest of the theory through the superpotential. The 2 quarks $q_m$ can be rotated into each other, thereby endowing the Lagrangian with an $SU(2)_b$ flavor symmetry. The matter content of the Lagrangian and its \underline{classical} symmetries can be summarized as in \eqref{eq:A1A3}.
\be
	\centering
	\begin{tabular}{|c|c|c|c|c|c|c|}
		\hline
		fields & $SU(2)_{\text{color}}$& $SU(2)_{b}$ & $U(1)_q$ & $U(1)_T$ & $U(1)_R$ & $ U(1)_T - \frac{3}{2} U(1)_q$\\
		\hline \hline
		$q$ & $\mathbf{2}$ & $\mathbf{2}$ & 1& $-\frac{1}{2}$ & $r_q$ & $-2$\\ \hline
		$\phi$ & {\rm adj} & $\mathbf{1}$ & 0 & 1 & $r_{\phi}$ & $1$ \\ \hline
		$M_{3}$ & $\mathbf{1}$ & $\mathbf{1}$ & $-2$ & 1 & $2- 2 r_q$ & $4$  \\ \hline
		$\beta$ & $\mathbf{1}$ & $\mathbf{1}$ & 0 & $-2$ & $2-2 r_\phi$ & $-2$\\ \hline
	\end{tabular} 
	\label{eq:A1A3}
\ee
The superpotential is given by
\be
W=M_3 {\rm Tr} qq + \beta {\rm Tr} \phi^2 \ .
\ee  
Note that, $U(1)_q$ and $U(1)_T$ are symmetries of the Lagrangian only at the classical level. In 4d they are both independently anomalous with only the linear combination given by $U(1)_A = U(1)_T - \frac{3}{2} U(1)_q $ being non-anomalous. Since, there are no anomalies in 3d, both $U(1)_q$ and $U(1)_T$ will survive as the symmetries of the CFT obtained at the fixed point of the dimensionally reduced Lagrangian. 

Requiring the 4d R-symmetry to be non-anomalous enforces the following constraint on the R-charges:
\be
r_q + 2 r_\phi = 1 \ .
\ee 
Thus in 4d the R-symmetry of the IR-fixed point belongs to a one-parameter family, the parameter being fixed by a-maximization \cite{Intriligator:2003jj}. Upon a-maximizing, we find that the 4d R-charges are given by 
\be
r_q= \frac{5}{9}, r_\phi = \frac{2}{9}, \ r_{M_3} = \frac{8}{9}, \ r_\beta = \frac{14}{9} \ .
\ee 

When considering the dimensional reduction of a 4d theory to 3d, one needs to check if a monopole superpotential can be generated or not. In the case at hand, Benvenuti and Giacomelli argued that a monopole superpotential will not be generated. This, coupled with the fact that there are no anomalies in 3d, implies that the 3d IR R-symmetry belongs to a two-parameter family which can be chosen to be $\{r_q, r_\phi\}$. These are fixed by extremizing the $S^3$ partition function of the theory. Benvenuti and Giacomelli showed that the corresponding values of R-charges are
\be
r_q = \half,\ r_\phi = 0, \ r_{M_3} = 1 , \ r_\beta=2 \ .
\ee  

\subsection{The 4d chiral ring} \label{sec:A1A34dchiral}
Let us now consider the 4d chiral ring of the $(A_1,A_3)$ Lagrangian. 
Before imposing the F-term conditions following from the superpotential, the ring of gauge invariant operators that can be formed from the chiral superfields listed in \eqref{eq:A1A3} is generated by the all possible monomials formed from the product over the following letters \footnote{Here $\a, \b$ are indices labeling the components of an $SU(2)_{\rm color}$-doublet while $m,n$ label the components of an $SU(2)_b$-doublet.}: 
\be
{\rm Tr} qq := \epsilon_{m n}\epsilon^{\a \b}q^m_{\a} q^n_{\b}, \ {\rm Tr}\phi^2, \ M_3 , \ \beta , \mu^{(m,n)}_l:= q_{\a}^m (\phi^l)^{\a \b} q_{\b}^n \ .
\ee 
Now, the equation of motion of $M_3$ throws ${\rm Tr} qq$ out of the chiral ring while the equation of motion of $\beta$ forces ${\rm Tr}\phi^2 = 0$. By the identities
\be
\phi^{2n} = (\frac{1}{2}{\rm Tr}\phi^2)^n \mathds{1} \ \text{and} \ \phi^{2n+1} = (\frac{1}{2}{\rm Tr}\phi^2)^n \phi \ , \forall n \geq 1 \ ,
\label{eq:SU2AdjFieldRel}
\ee 
obeyed by all $\phi \in \mathfrak{su}(2)$, it then follows that $\mu^{(a,b)}_n = 0, \  \forall n \geq 2$. The operator $\beta$ is quantum mechanically removed from the chiral ring. This is because a non-zero vev for $\beta$ lands us on a theory with no SUSY vacua. Alternately, it can be shown that $\beta$ is $Q$-exact with respect to the extra supercharges that emerge at the IR fixed point and hence $\beta$ is not in the chiral ring. In summary, the above arguments imply that the 4d chiral ring is given by 
\be
\mu^{(m,n)}:= q_\a^m \phi^{\a\b} q_\b^n , \ \text{and} \  M_3 \ .
\ee 
As was argued in \cite{Maruyoshi:2016aim}, $M_3$ generates the Coulomb branch of the $(A_1,A_3)$ theory realized at the IR fixed point of the above Lagrangian. We notice that $\mu^{(a,b)}$ transforms in the adjoint irrep. of $SU(2)_b$ with its IR scaling dimensions being 2. It is therefore natural to identify $\mu^{(m,n)}$ as the moment map operator of $SU(2)_b$. Thus we see that $\mu^{(m,n)}$ generates the Higgs branch of $(A_1, A_3)$ theory. Note that ${\rm det} \mu = 0$, which is the algebraic relation needed to define $\mathbb{C}^2/\mathbb{Z}_2$. Once again this is consistent with the identification of $\mu^{(a,b)}$ as the generators of the Higgs branch of $(A_1,A_3)$. 

\subsection{The 3d chiral ring}
In 3d, the analysis of section \ref{sec:A1A34dchiral} can be applied without any change. Additionally, there are two more chiral ring generators corresponding to the $SU(2)_{\rm color}$ -  monopole operator $\mathfrak{M}$ and the dressed monopole operator $\{\mathfrak{M}\phi\}$. Thus the list of 3d chiral ring generators and their charges with respect to the various symmetries can be listed as in \eqref{eq:A1A33dchiral}.
\be
\centering
\begin{tabular}{|c|c|c|c|c|}
	\hline
	Chiral Op. & $U(1)_{T} \subset SU(2)_T$& $U(1)_b \subset SU(2)_{b}$ & $U(1)_q$ & $U(1)_R$\\
	\hline \hline
	$\mu^{11}$ & $0$ & $+2$ &$+2$ & $+1$ \\ \hline
	$\mu^{12}$ & $0$ & $0$ &$+2$ & $+1$ \\ \hline
	$\mu^{22}$ & $0$ & $-2$ &$+2$ & $+1$ \\ \hline \hline
	$M_3$ & $1$ &$0$& $-2$ & 1 \\ \hline
	$\{\mathfrak{M}\phi \}$ & 0 & $0$ & $-2$ & 1 \\ \hline
	$\mathfrak{M}$ & $-1$ &$0$& $-2$ & 1 \\ \hline
\end{tabular}
\label{eq:A1A33dchiral}
\ee 
Here the charges of the monopole operator have been determined by using the formula given in \cite{Benini:2011cma}.

By comparing the entries in \eqref{tab:TSU2chiralOp} and \eqref{eq:A1A33dchiral}, one can easily see that the 3d chiral ring of the $(A_1,A_3)$ Lagrangian matches with that of the $T[SU(2)]$ theory. The correspondence between the chiral generators being given by 
\be
\centering
\begin{tabular}{|c|c|}
	\hline
	$T[SU(2)]$ & $(A_1,A_3)$ Lagrangian\\
	\hline \hline
	$P_1 \widetilde{P}_2$ &$\mu^{11}$  \\ \hline
	$\ P_1 \widetilde{P}_1 - P_2 \widetilde{P}_2$ &$\mu^{12}$  \\ \hline
	$\widetilde{P}_1 P_2$ &$\mu^{22}$  \\ \hline \hline
	$\mathfrak{M}^+$ & $M_3$  \\ \hline
	$\phi$ & $\{\mathfrak{M}\phi \}$  \\ \hline
	$\mathfrak{M}^-$ &$\mathfrak{M}$  \\ \hline
\end{tabular}
\label{eq:A1A3TSU2Match}
\ee
It will be useful to note that the gauge singlet field $M_3$ of the $(A_1,A_3)$ Lagrangian maps to the monopole operator $\mathfrak{M}^+$ of the $T[SU(2)]$ theory.

\subsection{3d Superconformal Index}
The superconformal index of the 3d $(A_1,A_3)$ Lagrangian can be written as the following integral
\be
I_{(A_1,A_3)} = \sum_{m \in \mathbb{Z}_{\geq 0}}\oint \frac{d z}{2 \pi i z} \frac{1}{|\mathcal{W}(m)|} Z_{\rm vec} (x,b,v,z) Z_{\phi}(x,b,v,z) Z_q(x,b,v,z) Z_{M_3} (x,b,v,z) Z_\beta(x,b,v,z)  \nonumber \ , \\
\ee 
where
\be
|\mathcal{W}(m)| &=& 1 + \delta_{m,0} \ , \\
Z_{\rm vec} (x,b,v,z) &= & x^{-2m} (1-z^2 x^{2m})(1-z^{-2} x^{2m}) \ , \\
Z_{\phi}(x,b,v,z) &= & (x t^{-1})^{2m} \times \nonumber \\ &~~~~& \prod_{k=0}^{\infty} \frac{1-z^{-2} t^{-1} x^{2k + 2m +2} }{1-z^{2} t x^{2k + 2m }} \frac{1- t^{-1} x^{2k +2} }{1- t x^{2k }}\frac{1-z^{2} t^{-1} x^{2k + 2m +2} }{1-z^{-2} t x^{2k + 2m }} \ ,  \\
Z_q (x,b,v,z) &=& (x t v^{-2})^{m} \times \nonumber \\ & &  \prod_{\sigma_1,\sigma_2 \in \{\pm\}}\prod_{k=0}^{\infty} \frac{1-z^{\sigma_1} b^{\sigma_2} t^{\half} v^{-1} x^{2k + m + \frac{3}{2}} }{1-z^{-\sigma_1} b^{-\sigma_2}t^{-\half} v x^{2k + m + \half }} \ , \\
Z_{M_3}(x,b,v,z) &= & \prod_{k=0}^{\infty} \frac{1- t^{-1} v^{2} x^{2k+1} }{1- t v^{-2} x^{2k + 1 }} \ \text{and} \\
Z_{\beta}(x,b,v,z) &= & \prod_{k=0}^{\infty} \frac{1- t^{2} x^{2k} }{1-  t^{-2} x^{2k + 2 }} \ .
\ee 
Upon explicit evaluation, we find that the superconformal index matches exactly with that of the $T[SU(2)]$ theory i.e. the series given in \eqref{eq:TSU2Index}.

One can also consider the $S^3$ partition function of the $(A_1,A_3)$ Lagrangian and show that it matches with that of the $T[SU(2)]$ theory, as was done numerically in \cite{Benvenuti:2017lle} and analytically in \cite{Aghaei:2017xqe}.

\section{The $(A_1,D_3)$ Lagrangian}
\label{sec:A1D3}
Let us now move on to the $(A_1,D_3)$ Lagrangian. 
Its field content along with the gauge and classical flavor symmetries are summarized in \eqref{tab:A1D3}.
\be
	\centering
	\begin{tabular}{|c|c|c|c|c|c|c|}
		\hline
		fields & $SU(2)_{\text{color}}$& $SO(3)_{b}$ & $U(1)_T$ & $U(1)_q$ & $U(1)_R$ & $U(1)_T-\frac{3}{2} U(1)_q$\\
		\hline \hline
		$q_1$ & $\mathbf{2}$ & $\mathbf{3}$ & $\frac{1}{4}$& $\frac{1}{2}$ & $1-\frac{r_\phi}{2}$ & $-\frac{1}{2}$ \\ \hline
		$q_2$ & $\mathbf{2}$ & $\mathbf{1}$ & $-\frac{1}{4}$ & $\frac{3}{2}$ & $1-\frac{r_{M_3} + r_\phi}{2}$ & $-\frac{5}{2}$ \\ \hline
		$\phi$ & {\rm adj} & $\mathbf{1}$ & $-\half$ & $-1$ & $r_\phi$ & $1$ \\ \hline
		$M_{3}$ & $\mathbf{1}$ & $\mathbf{1}$ & 1 & $-2$ & $r_{M_3}$ &$4$ \\ \hline
		$\beta$ & $\mathbf{1}$ & $\mathbf{1}$ & 1 & $2$ & $2- 2 r_\phi$ & $-2$ \\ \hline
	\end{tabular}
	\label{tab:A1D3}
\ee
Note that for the time being we have included the flipping field $\beta$ in our Lagrangian. We will soon show that it must not be included in the dimensionally reduced Lagrangian.

The $4d$ superpotential is therefore given by
\be
W={\rm Tr}q_1\phi q_1 + M_3 {\rm Tr}q_2 \phi q_2 + \beta ~ {\rm Tr}\phi^2  \ .
\label{eq:A1D3SuperPot}
\ee 
In 4d, $U(1)_T$ and $U(1)_q$ are both anomalous with $U(1)_{\widetilde{A}} = U(1)_T -\frac{3}{2} U(1)_q$ being non-anomalous. Thus only $U(1)_{\widetilde{A}}$ is a symmetry of the 4d quantum mechanical theory. On the contrary, in 3d there are no anomalies and hence both $U(1)_T$ and $U(1)_q$ will be symmetries of the quantum mechanical Lagrangian. 

Requiring the 4d R-charge to be non-anomalous gives us the following constraint
\be
r_{M_3} = 4 r_\phi \ .
\ee 
Thus in 4d, the IR-symmetry a priori belongs to a one parameter family that can be parametrized by $r_{M_3}$. This is fixed by a-maximization with the result being 
\be
r_{M_3} = \frac{8}{9} , \ r_\phi = \frac{2}{9}, \ r_{q_1} = \frac{8}{9}, \ r_{q_2} = \frac{4}{9} \ \text{and} \ r_{\beta} = \frac{14}{9} \ .
\ee 
Using these to compute the central charges $a$ and $c$, it can be checked that they match with the $(A_1,D_3)$ AD theory. Moreover, by comparing the respective quantum numbers, it can be checked that $M_3$ maps to the Coulomb branch operator of the $(A_1,D_3)$ theory \cite{Maruyoshi:2016aim, Agarwal:2016pjo}.


\subsection{The 4d chiral ring}
We will now compute the 4d chiral ring of this Lagrangian. 
The equation of motion of $q_1$ implies that $(\phi q_1)^{a \alpha}=0$ \footnote{Here $\alpha,\beta$ are indices labeling the components of an $SU(2)_{\rm color}$ doublet while $a,b$ label the components in the vector representation of $SO(3)_{\rm flavor}$.  The indices $i,j$ label the two kinds of $SU(2)_{\rm color}$ : $q_1$ and $q_2$, in \eqref{tab:A1D3}. }. 
It therefore follows that $(q_1)^{a}_{\alpha} \phi^{\alpha \beta} (q_1)^{b}_{\beta} $ and $(q_2)_{\alpha} \phi^{\alpha \beta} (q_1)^{a}_{\beta} $ are trivial in the chiral ring. 
At the same time the equation of motion of $M_3$ throws $(q_2)_{\alpha} \phi^{\alpha \beta} (q_2)_{\beta} $ out of the chiral ring. This also implies that all operators of form ${\rm Tr} q_2 \phi^{(2l+1)} q_2 $ are trivial in the chiral ring since they can be factorized as ${\rm Tr} q_2 \phi^{(2l+1)} q_2 =  ({\rm Tr} q_2 \phi q_2 ) (\half {\rm Tr\phi^2})^l =0  $ upon using the identities of \eqref{eq:SU2AdjFieldRel}. The operators ${\rm Tr} q_2 \phi^{(2l)} q_2 $ are trivially zero, as can be easily seen by applying  \eqref{eq:SU2AdjFieldRel}.
Hence
\be
(q_i)^{a}_{\alpha} (\phi^l)^{\alpha \beta} (q_j)^{b}_{\beta}=0 \ , \forall l\geq 1 \ \& \ i,j \in \{1,2\}
\label{eq:A1D34dCh}
\ee

The equation of motion of $\beta$ gives ${\rm Tr}\phi^2 =0$, hence ${\rm Tr}\phi^2$ is no longer in the chiral ring.  

In the cases considered in \cite{Benvenuti:2017lle,Benvenuti:2017kud}, $\beta$ was quantum mechanically removed from the chiral ring. This is because a non-zero vev for $\beta$ lands us on a theory with no SUSY vacua. However, in the present case, giving vev to $\beta$ and upon integrating out $\phi$ gives us a 4d $\mathcal{N}=1$ $SU(2)$ gauge theory with 4 fundamental chiral doublets. This has a quantum deformed moduli space. The existence of valid SUSY vacuum moduli space then implies that we can not apply the same reasoning as \cite{Benvenuti:2017lle,Benvenuti:2017kud} to claim that $\beta$ is not part of the chiral ring. However, an independent argument to claim that $\beta$ is not part of the chiral ring is that $\beta$ is $Q$-exact with respect to the accidental supercharge which emerges at the IR fixed point. To see that this is indeed the case, notice that in 4d $\mathcal{N}=2$ theories, any Coulomb branch operator $u$ lives in a multiplet which contains another scalar $v$, given by
\be
\label{eq:colbrdec}
v = \int d^2 \widetilde{\theta} u \ ,
\ee 
where $\widetilde{\theta}$ is the Grassman parameter of the hidden supercharge at the IR fixed point. It therefore follows that 
\be
\Delta_v = \Delta_u + 1 \ , \text{and} \ R_v = R_u + \frac{2}{3} \ .
\ee 
Clearly $v$ is $Q$-exact and will not be a part of the chiral ring. We now notice that in the case at hand, the singlet field $M_3$ has R-charge $\frac{8}{9}$ and hence its counterpart, as defined above, will have R-charge $\frac{14}{9}$. This matches exactly with the R-charge of $\beta$ and hence we claim that $\beta$ is related to $M_3$ through the relation \eqref{eq:colbrdec}. This implies that $\beta$ is $Q$-exact i.e. it is trivial in the chiral ring.

A priori, it appears that the chiral ring generators are : $\{ M_3, \ \epsilon^{\alpha \beta}(q_1)^{a}_{\alpha} (q_1)^{b}_{\beta} , \ \epsilon^{\alpha \beta}(q_1)^{a}_{\alpha} (q_2)_{\beta}  \}$. Their charges and scaling dimensions under various global symmetries are listed in  \eqref{tab:A1D3ch}. 
\be 
	\centering
	\begin{tabular}{|c|c|c|c|c|}
		\hline
		chiral op. & $SO(3)_{b}$ & $\widetilde{A}$ & $R_{IR}$ & $\Delta$ \\
		\hline \hline
		$M_3$ & $\mathbf{1}$ & 4 & $\frac{8}{9}$ & $\frac{4}{3}$ \\ \hline
		$\epsilon^{\alpha \beta}(q_1)^{a}_{\alpha} (q_1)^{b}_{\beta}$ & $\mathbf{3}$ & -1 &$\frac{16}{9}$ & $\frac{8}{3}$ \\ \hline
		$\epsilon^{\alpha \beta}(q_1)^{a}_{\alpha} (q_2)_{\beta}$ & $\mathbf{3}$ & -3 & $\frac{4}{3}$ & $2$ \\ \hline
	\end{tabular}
	\label{tab:A1D3ch}
\ee 
It was already established in \cite{Agarwal:2016pjo} that $M_3$ is isomorphic to the Coulomb branch generator of the $(A_1,D_3)$ AD-theory. From the analysis presented here, it is clear that $\epsilon^{\alpha \beta}(q_1)^{a}_{\alpha} (q_2)_{\beta}$ has the right quantum numbers to be identified with the moment map operator of the $SU(2)_{\rm flavor} \simeq SO(3)_{b}$ symmetry enjoyed by the $(A_1,D_3)$ AD-theory. However, the $(A_1,D_3)$ AD-theory has no other independent chiral operators. We therefore claim that quantum mechanical corrections to the chiral ring will cause the operator  $ \epsilon^{\alpha \beta}(q_1)^{a}_{\alpha} (q_1)^{b}_{\beta} =0 $. That this is the case can also be tested by considering the 4d superconformal index of our theory. The general 4d $\nn{2}$ index can be defined as
\be
\mathcal{I}_{\nn{2}}(\mathfrak{t},y,v) = {\rm Tr}(-1)^F \mathfrak{t}^{2(E+j_2)}y^{2 j_1}v^{-I_3 + \frac{r}{2}} \ ,
\ee 
with $E$ being the scaling dimension of the operator that contributes to the index, $(j_1,j_2)$ are its quantum numbers under the $SO(3,1) \simeq SU(2)_1 \times SU(2)_2$ Lorentz symmetry and $(I_3, r)$ are its $\nn{2}$ R-charges. It therefore follows that the operator $\epsilon^{\alpha \beta}(q_1)^{a}_{\alpha} (q_1)^{b}_{\beta}$, if present, should contribute  a term of form $\mathfrak{t}^{\frac{16}{3}} v^{-\frac{1}{3}} \chi _3^b$ , with its conformal descendants contributing terms which are of higher order in $\mathfrak{t}$. The 4d superconformal index of the $(A_1,D_3)$ Lagrangian was explicitly computed in \cite{Agarwal:2016pjo}. We reproduce it here
\be
\CI_{\CN=2}^{(A_1, D_{3})} &=1+\mathfrak{t} ^{8/3} v^{4/3}-\mathfrak{t} ^{11/3} v^{1/3}\chi_2^y+\mathfrak{t} ^4 v^{-1} \chi_{3}^b+\mathfrak{t} ^{14/3}v^{-2/3} 
+\mathfrak{t} ^{16/3} v^{8/3}  + \mathcal{O} (\mathfrak{t} ^{17/3}) \nonumber
\ee 
The absence of $\mathfrak{t}^{\frac{16}{3}} v^{-\frac{1}{3}} \chi _3^b$  therefore confirms our expectation. However, it will be nice to better understand the physical mechanism that kills this operator. 

\subsection{The 3d chiral ring}
\label{sec:A1D33dCh}

Taking cue from \cite{Benvenuti:2017kud}, we will assume that the absence of $\beta$ and $\epsilon^{\alpha \beta}(q_1)^{a}_{\alpha} (q_1)^{b}_{\beta}$ from the 4d chiral ring will continue to be true in 3d also. This implies that no monopole superpotential can be generated \footnote{Note that the assumption about $\beta$ not being part of the 3d chiral ring is somewhat ad hoc in nature. One way to see this is to notice that in 3d, $\beta$ can be safely given a non-zero vev without causing a quantum mechanical break-down of supersymmetry.  We will therefore loosen this assumption and consider the consequences of including a monopole superpotential in section \ref{sec:MonSup}.} and therefore
in addition to the generators of the 4d chiral ring presented in the previous section, the 3d chiral ring will have 2 additional generators : the basic $SU(2)_{\rm color}$ - monopole operator $\mathfrak{M}$ and the dressed monopole operator $\{\mathfrak{M}\phi\}$. 

 Let us now consider the match between the 3d chiral ring of the $(A_1,D_3)$ Lagrangian and that of the $T[SU(2)]$ theory. Since the operator $M_3$ corresponds to the generator of 4d IR Coulomb branch in both $(A_1,A_3)$ and $(A_1,D_3)$ Lagrangian, thus we expect that upon 3d reduction, the operator $M_3$ of the $(A_1,D_3)$ matches with $\mathfrak{M}^+$ of $T[SU(2)]$, as this was also the case in the 3d reduction of the $(A_1,A_3)$ Lagrangian. Hence we must normalize $U(1)_T$ of $(A_1,D_3)$ such that it assigns charge $+1$ to $M_3$. Similarly, we must normalize $U(1)_q$ of $(A_1,D_3)$ to be such that it assigns charge $-2$ to $M_3$. 
 The normalization of $U(1)_T$ and $U(1)_q$ of \eqref{tab:A1D3} were chosen exactly in this way. 

Also, it must be that the moment map operator of the $SU(2)_b \simeq SO(3)_b$ flavor symmetry matches across the 3 Lagrangians. This implies that the operator $\epsilon^{\alpha \beta}(q_1)^{a}_{\alpha} (q_2)_{\beta}$ must map to the $SU(2)_b$ moment map $\mu^{(a,b)}$ of $T[SU(2)]$. This forces us to normalize $U(1)_T$ and $U(1)_q$ symmetries of $(A_1,D_3)$ in such a way that the $U(1)_T$ charge of $q_1$ and $q_2$ adds up to $0$ while their $U(1)_q$ charge must add up to $2$. This was the additional constraint which helped us assign $U(1)_T$ and $U(1)_q$ charges to all the fields in \eqref{tab:A1D3}.

The above two maps can also be used to compute the expected 3d R-symmetry at the IR fixed point of the $(A_1,D_3)$ theory. Thus we want the R-charges to be such that $r_{M_3} = r_{q_1} + r_{q_2} = 1$. This implies that we expect that at the IR-fixed point the R-charges of the various fields in the Lagrangian are 
\be
r_{M_3} = 1 , \ r_\phi = \frac{1}{2}, \ r_{q_1} = \frac{3}{4}, \ r_{q_2} = \frac{1}{4} \ \text{and} \ r_{\beta} = 1 \ .
\label{eq:A1D3R3d}
\ee 
The R-charge of the monopole operator $\mathfrak{M}$ is given by
\be
\nonumber r_{\mathfrak{M}} 
& =& 4- 2 r_{\phi} - 3 r_{q_1} -  r_{q_2} \nonumber \\
&=& \frac{r_{M_3}}{2} \ .
\ee 
Upon substituting from \eqref{eq:A1D3R3d}, we then find that 
\be
r_{\mathfrak{M}}=\half \ \text{and} \ r_{\{\mathfrak{M}\phi \}} = 1 \ .
\ee 
Similarly we can compute the $U(1)_T$ and $U(1)_q$ charges of $\mathfrak{M}$ and find that 
\be
T_{\mathfrak{M}} = \half , Q_{\mathfrak{M}_3} = -1, \ T_{\{\mathfrak{M}\phi \}}  = 0, \ Q_{\{\mathfrak{M}\phi \}} = -2 \ .
\ee
The R-charge of $\mathfrak{M}$ being $\half$ indicates that it decouples as a free field in the IR. Hence forth, when we talk about the 3d SCFT described by the $(A_1,D_3)$ Lagrangian, we mean the interacting sector obtained after decoupling $\mathfrak{M}$.

We almost have the full dictionary between the 3d chiral ring of interacting sector in the IR of $(A_1,D_3)$ Lagrangian and $T[SU(2)]$. However, there seems to be no chiral operator in the IR SCFT of $(A_1,D_3)$ Lagrangian that maps to $\mathfrak{M}^-$ of $T[SU(2)]$. The way out seems to be from observing that if we omit the term $\beta {\rm Tr}\phi^2$ from the 3d superpotential (for the time being in an ad hoc manner; we can still include this term in the 4d superpotential), then the 3d chiral ring will also contain ${\rm Tr}\phi^2$ which has just the right charges to match with $\mathfrak{M}^-$ of $T[SU(2)]$. Note that, in deleting this term from 3d superpotential, we will end up decoupling the field $\beta$ from the 3d theory. 
We will soon verify our claim by computing the $S^3$ partition function and the superconformal index of the dimensionally reduced $(A_1,D_3)$ Lagrangian.  
Before moving on, we also wish to point out that excluding the flipping field $\beta$ from the 3d Lagrangian will not change our analysis of the 3d chiral ring as we did not need to use the equation of motion of $\beta$ in  arriving at \eqref{eq:A1D34dCh} to establish the triviality of operators ${\rm Tr}(q_i)^{a} (\phi^l) (q_j)^{b}$.


\subsection{3d Superconformal Index}
 Let us consider the superconformal index of 3d $(A_1,D_3)$ Lagrangian sans the gauge singlet field $\beta$. All the fields will be assigned R-charges according to those listed in \eqref{eq:A1D3R3d}. 
 Upon accounting for the decoupling of $\mathfrak{M}$.
 the superconformal index of the $(A_1,D_3)$ Lagrangian can be written as 
\be
\label{eq:A1D3ind}
I_{(A_1,D_3)} &=& \frac{1}{Z_{\mathfrak{M}}(x,b,v)} \times \nonumber \\
& &\sum_{m \in \mathbb{Z}_{\geq 0}}\oint \frac{d z}{2 \pi i z} \frac{1}{\mathcal{W}(m)} Z_{\rm vec} (x,b,v,z) Z_{\phi}(x,b,v,z) Z_{q_1}(x,b,v,z) Z_{q_2}(x,b,v,z)Z_{M_3} (x,b,v,z)  \nonumber \ , \\
\ee 
where
\be
\mathcal{W}(m) &=& 1 + \delta_{m,0} \ , \\
Z_{\rm vec} (x,b,v,z) &= & x^{-2m} (1-z^2 x^{2m})(1-z^{-2} x^{2m}) \ , \\
Z_{\phi}(x,b,v,z) &= & (x t v^2)^{m} \times \nonumber \\ &~~~~& \prod_{k=0}^{\infty} \frac{1-z^{\pm2} t^{\half} v x^{2k + 2m +\frac{3}{2}} }{1-z^{\mp 2} t^{-\half} v^{-1} x^{2k + 2m + \half }} \frac{1- t^{\half} v x^{2k +\frac{3}{2}} }{1- t^{-\half} v^{-1} x^{2k+\half }} \ ,   \\
Z_{q_1} (x,b,v,z) &=& (x^{\frac{3}{4}} t^{-\frac{3}{4}} v^{-\frac{3}{2}})^{m} \times \nonumber \\ & &  \prod_{\sigma_1, \in \{\pm\}}\prod_{\sigma_2, \in \{2,0,-2\}}\prod_{k=0}^{\infty} \frac{1-z^{\sigma_1} b^{\sigma_2} t^{-\frac{1}{4}} v^{-\half} x^{2k + m + \frac{5}{4}} }{1-z^{-\sigma_1} b^{-\sigma_2}t^{\frac{1}{4}} v^{\half} x^{2k + m + \frac{3}{4} }} \ , \\
Z_{q_2} (x,b,v,z) &=& (x^{\frac{3}{4}} t^{\frac{1}{4}} v^{-\frac{3}{2}})^{m} \times \nonumber \\ & &  \prod_{\sigma_1, \in \{\pm\}}\prod_{k=0}^{\infty} \frac{1-z^{\sigma_1}  t^{\frac{1}{4}} v^{-\frac{3}{2}} x^{2k + m + \frac{7}{4}} }{1-z^{-\sigma_1} t^{-\frac{1}{4}} v^{\frac{3}{2}} x^{2k + m + \frac{1}{4} }} \ , \\
Z_{M_3}(x,b,v,z) &= & \prod_{k=0}^{\infty} \frac{1- t^{-1} v^{2} x^{2k+1} }{1- t v^{-2} x^{2k + 1 }} \ \text{and} \\
Z_{\mathfrak{M}}(x,b,v) &= & \prod_{k=0}^{\infty} \frac{1- v t^{-\half} x^{2k+\frac{3}{2}} }{1-  v^{-1}t^{\half} x^{2k + \half }} \ .
\ee 
Upon evaluating \eqref{eq:A1D3ind} explicitly, we find that it matches with the series expansion of $T[SU(2)]$ superconformal index given in \eqref{eq:TSU2Index}.


We thus notice that the 3d reduction of the$(A_1,D_3)$ Lagrangian as described here gives us a new dual of the $T[SU(2)]$. It will be interesting to see if, for certain application, this description gives us any advantage over the $T[SU(2)]$ theory itself or its previously known duals such as those proposed in \cite{Teschner:2012em} (also see \cite{Gang:2012ff}, who checked its duality to $T[SU(2)]$ at the level of index) and \cite{Gang:2013sqa}.   

\subsection{The $S^3$ partition function }
\label{eq:S3partfun}

\begin{figure}[t]
	\centering
    \begin{subfigure}{5cm}
	\includegraphics[width=5cm , height=5cm]{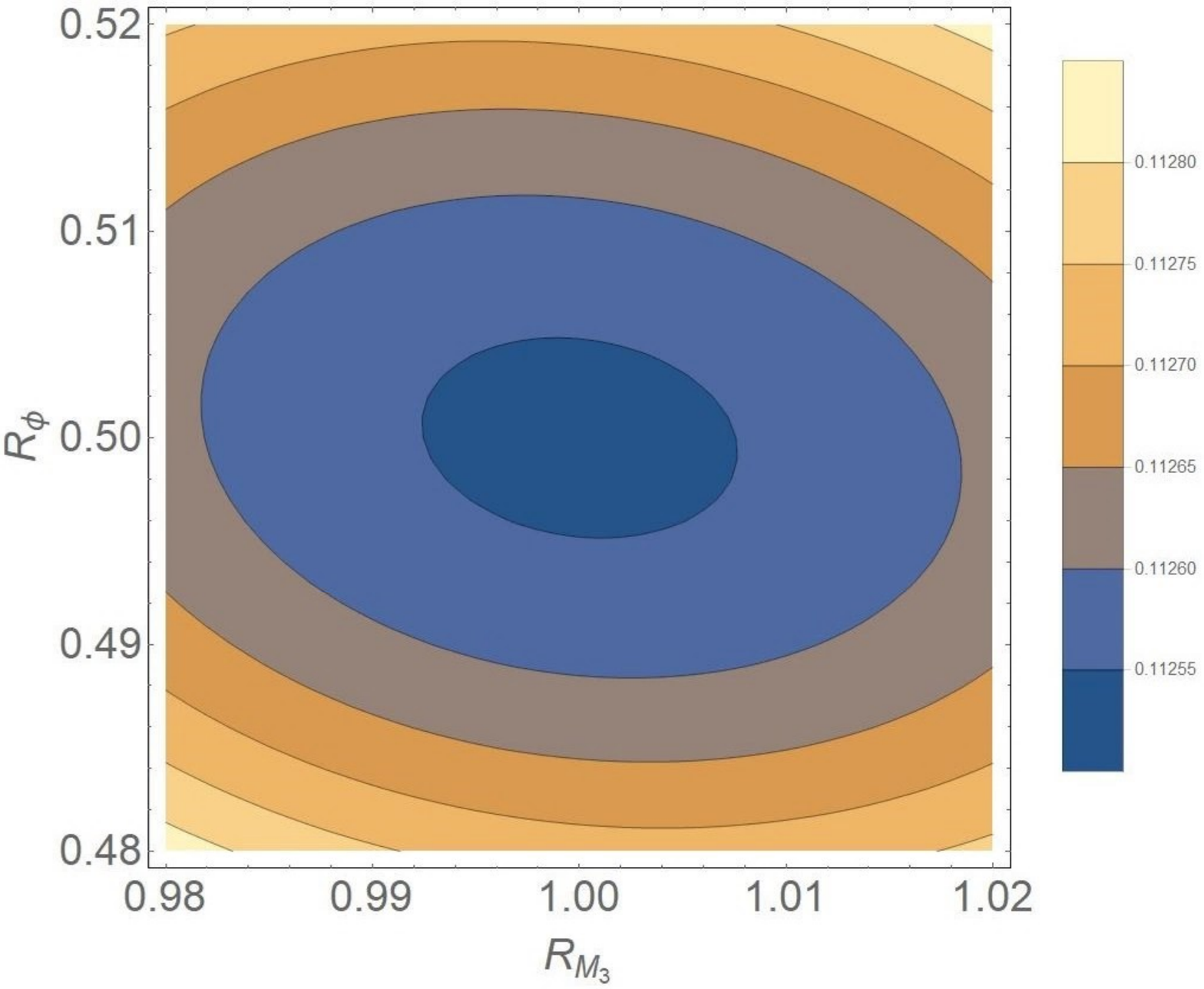}
	\caption{}
	\end{subfigure}
    \begin{subfigure}{5cm}
	\includegraphics[width=5cm , height=5cm]{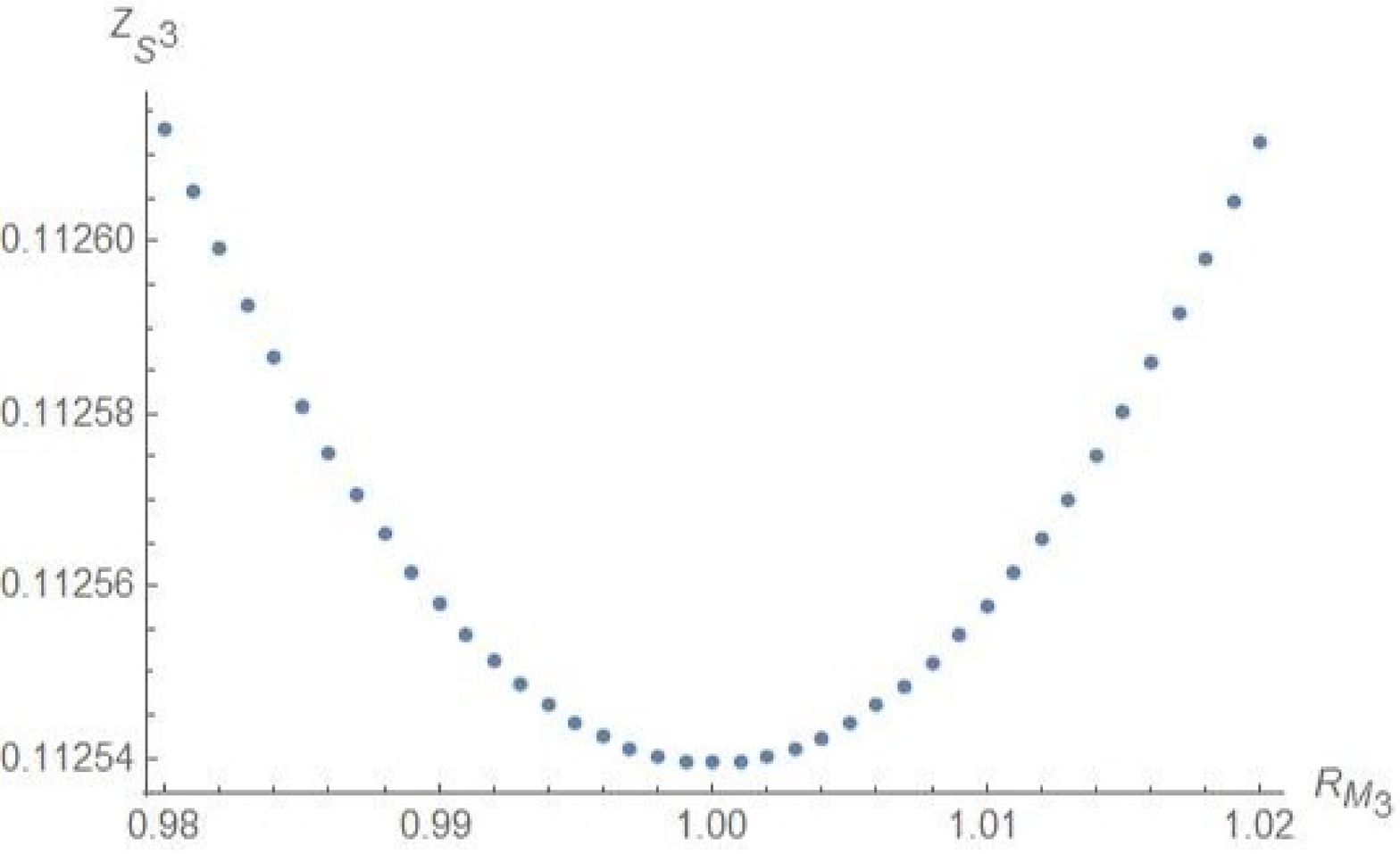}
	\caption{}
     \end{subfigure} 
     \begin{subfigure}{5cm}
	\includegraphics[width=5cm , height=5cm]{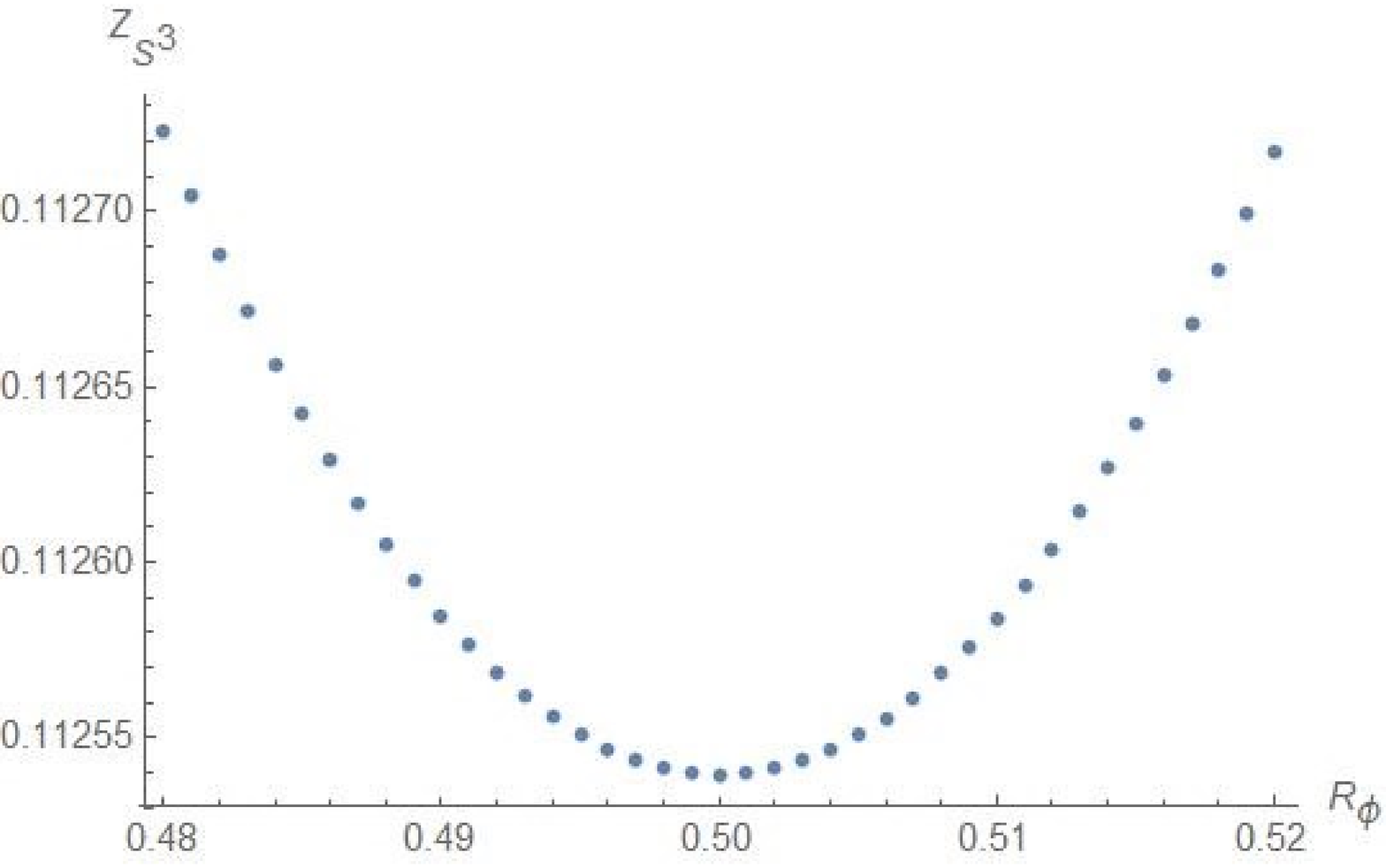}
	\caption{}
     \end{subfigure}
    \caption{Extremizing the partition function of the $(A_1,D_3)$ Lagrangian without the flipping field $\beta$. (a) Contour diagram of $\CZ_{S^3}^{(A_1,D_3)}$ vs. \ $\{ r_{M_3} , r_{\phi}\}$. (b) Plot of $\CZ_{S^3}^{(A_1,D_3)}$ vs. \ $r_{M_3}$ at $r_\phi=\half$. (c) Plot of $\CZ_{S^3}^{(A_1,D_3)}$ vs. \ $r_{\phi}$ at $r_{M_3}=1$.}
    \label{fig:3dplotsNobeta}
\end{figure}

\noindent {\textbf{The 3d $(A_1,D_3)$ theory without the flipping field $\beta$}} : Let us also compute the partition function of the 3d theory obtained by removing the $\beta {\rm Tr}\phi^2$ term in the superpotential. Recall that this implies that $\beta$ is no longer coupled to the 3d theory and therefore we don't include it's contribution to the partition function, which is then given by
\be
\mathcal{Z}_{S^3}^{(A_1,D_3)} &=& \frac{e^{l(1- r_{M_3})}}{2!} \times \nonumber \\ & &\int_{-\infty}^{\infty} \frac{dz}{2 \pi i z} (2 \sinh 2\pi z)^2  e^{l(1-r_{\phi} \pm  2 i z) + l(1-r_{\phi}) + 3 l(\frac{r_\phi}{2} \pm i z ) + l(\frac{r_\phi + r_{M_3}}{2} \pm i z ) } \nonumber \ . \\
\ee 
It is easy to check that the above integral attains its minima at $r_{M_3} = 1, \ r_{\phi} = \half$,  as can be seen from the plots shown in Figure \ref{fig:3dplotsNobeta}.
After factorizing out the contribution from the decoupled monopole i.e. dividing the partition function by $\frac{1}{\sqrt{2}}$, we find that at its minima i.e $r_{M_3} = 1, \ r_{\phi} = \half$, the numerical value of the partition function is approximately $\frac{1}{2 \pi}$ which matches with the 
numerical value of the $S^3$ partition function of $T[SU(2)]$.

\vspace{5pt}

\begin{figure}
	\centering
	\begin{subfigure}{5cm}
    \includegraphics[width=5cm]{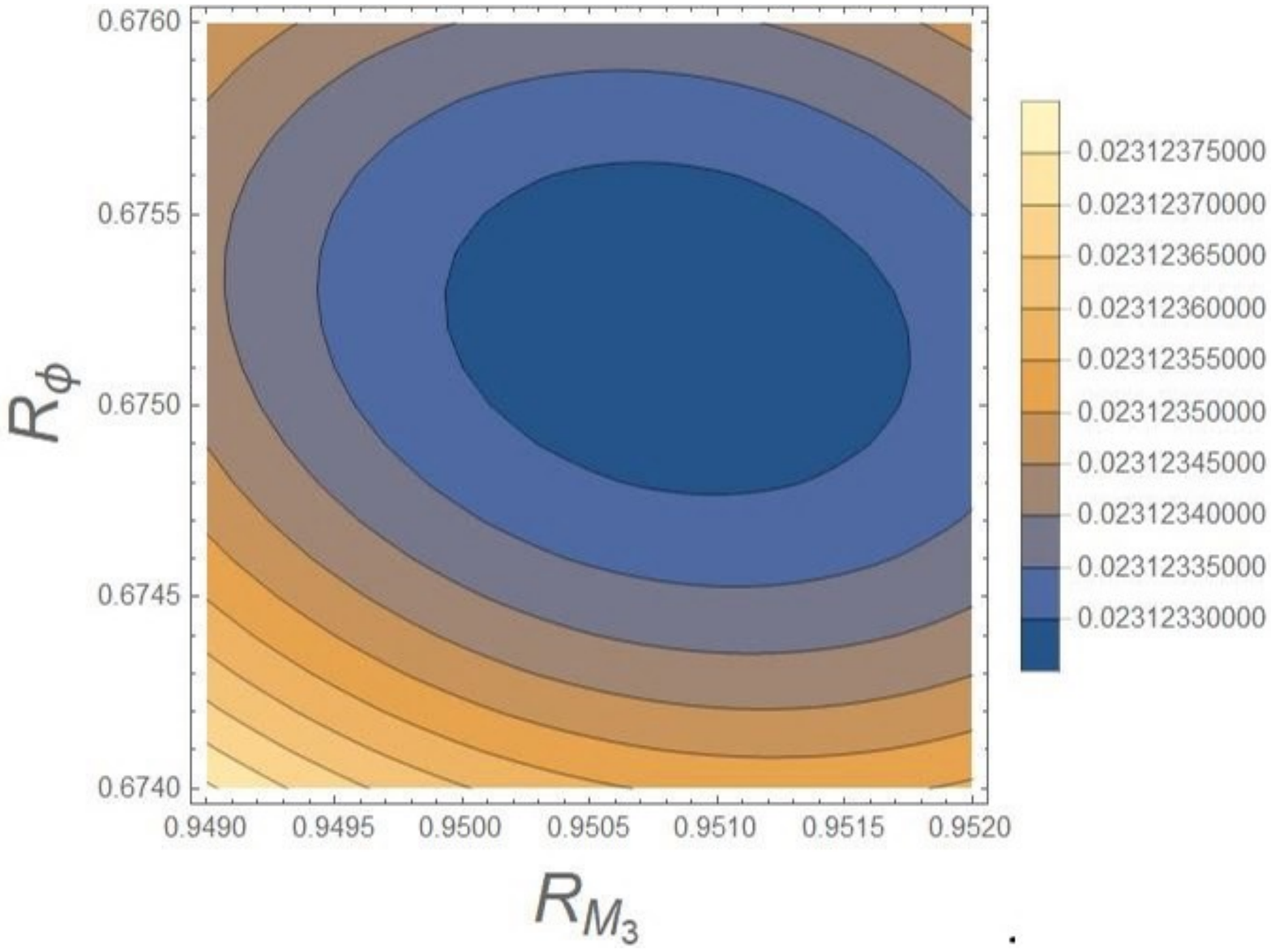}
	\caption{}
	\end{subfigure}
	\begin{subfigure}{5cm}
	\includegraphics[width=5cm]{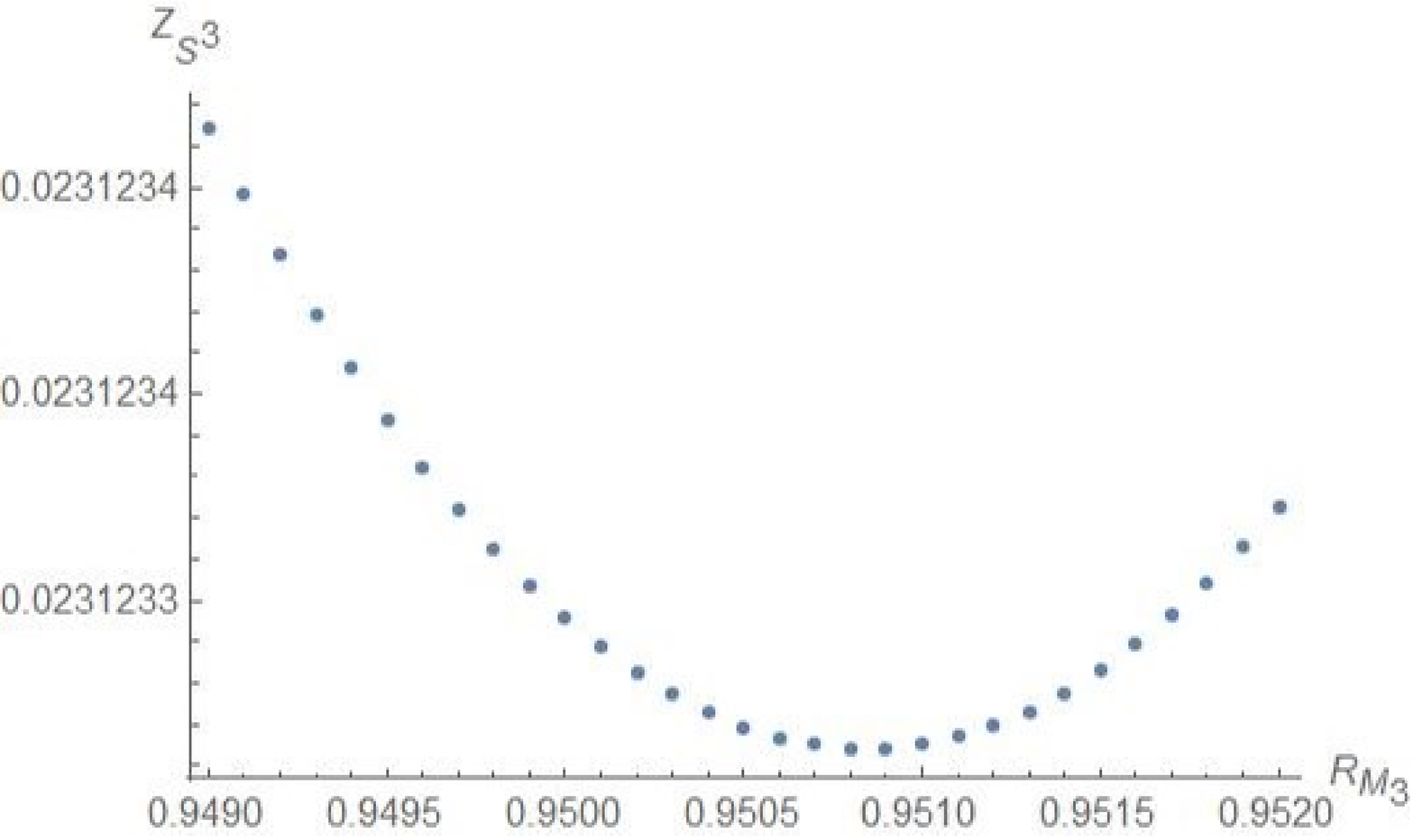}
	\caption{}
    \end{subfigure}
	\begin{subfigure}{5cm}
	\includegraphics[width=5cm]{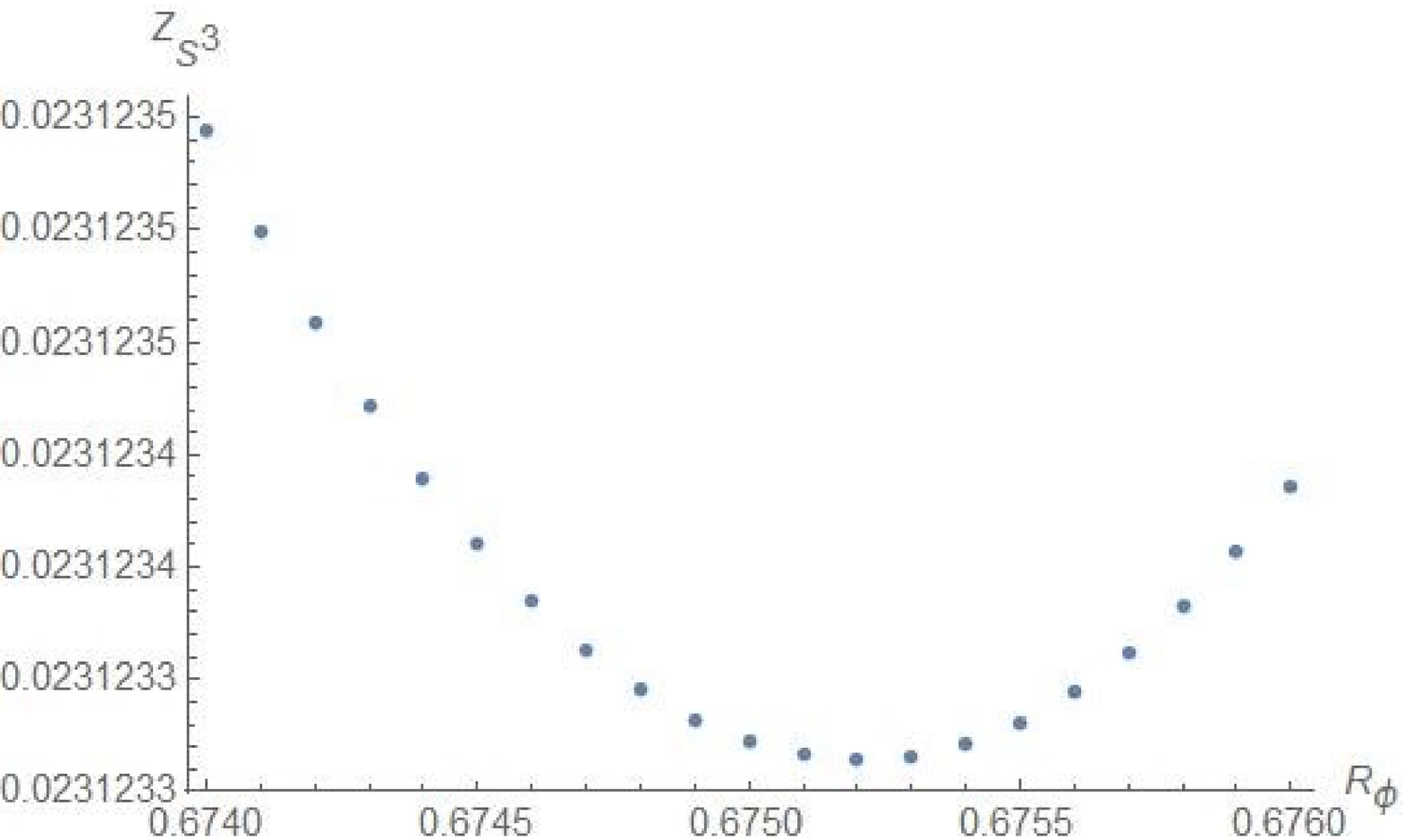}
	\caption{}
    \end{subfigure}	
	\caption{Extremizing the partition function of the $(A_1,D_3)$ Lagrangian including the flipping field $\beta$. The plots give the values before removing the contribution of the decoupled monopole operator. (a) Contour diagram of $\CZ_{S^3}^{(A_1,D_3)+ \beta}$ vs. \ $\{ r_{M_3} , r_{\phi}\}$. (b) Plot of $\CZ_{S^3}^{(A_1,D_3) + \beta}$ vs. \ $r_{M_3}$ at $r_\phi = 0.6752  $. (c) Plot of $\CZ_{S^3}^{(A_1,D_3)+ \beta}$ vs. \ $r_{\phi}$ at $r_{M_3}=0.9508$.   }
	\label{fig:BeforeMonDecop}
\end{figure}

\begin{figure}
		\centering
	\begin{subfigure}{5cm}
		\includegraphics[width=5cm]{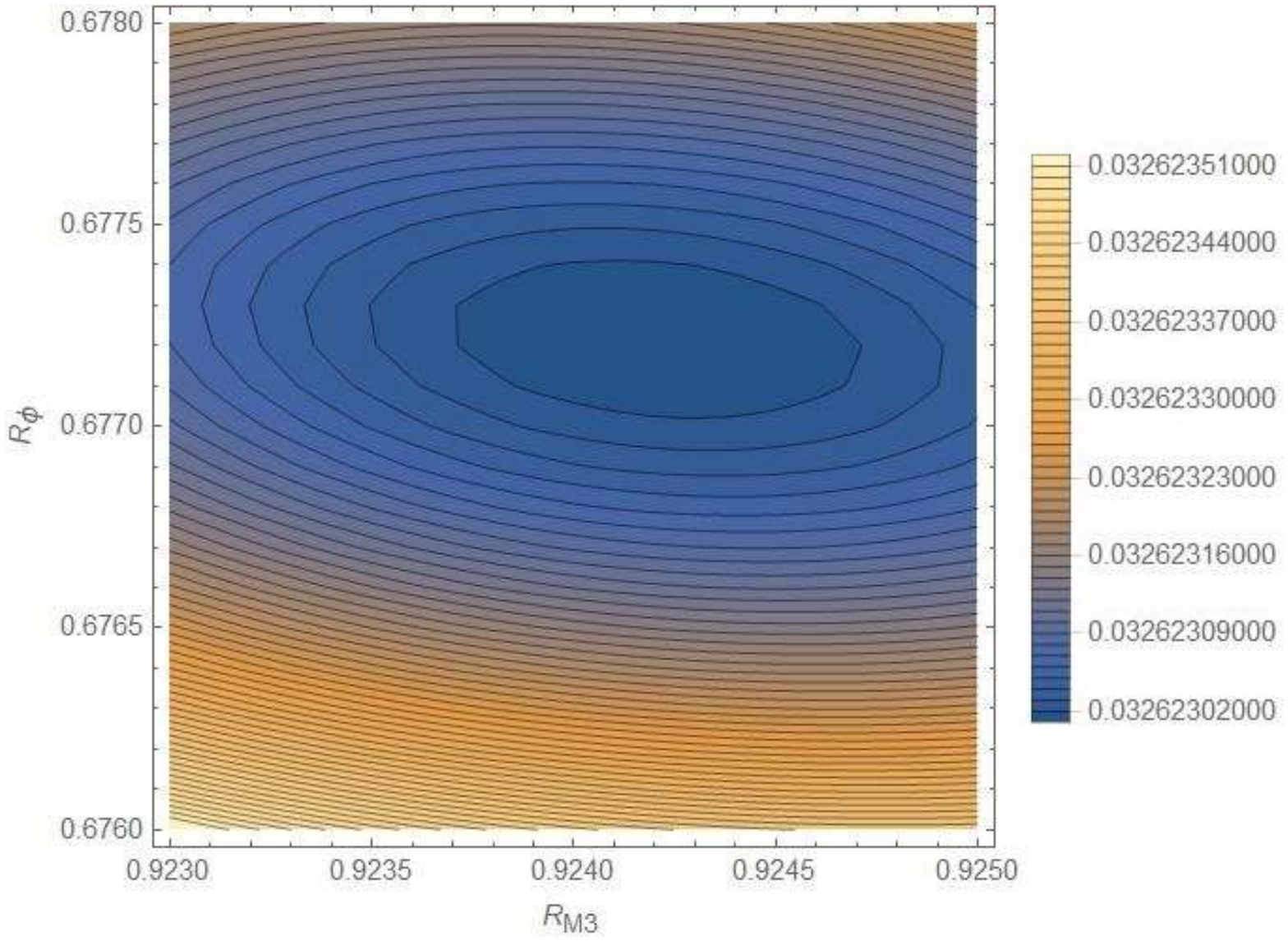}
		\caption{}
	\end{subfigure}
	\begin{subfigure}{5cm}
		\includegraphics[width=5cm]{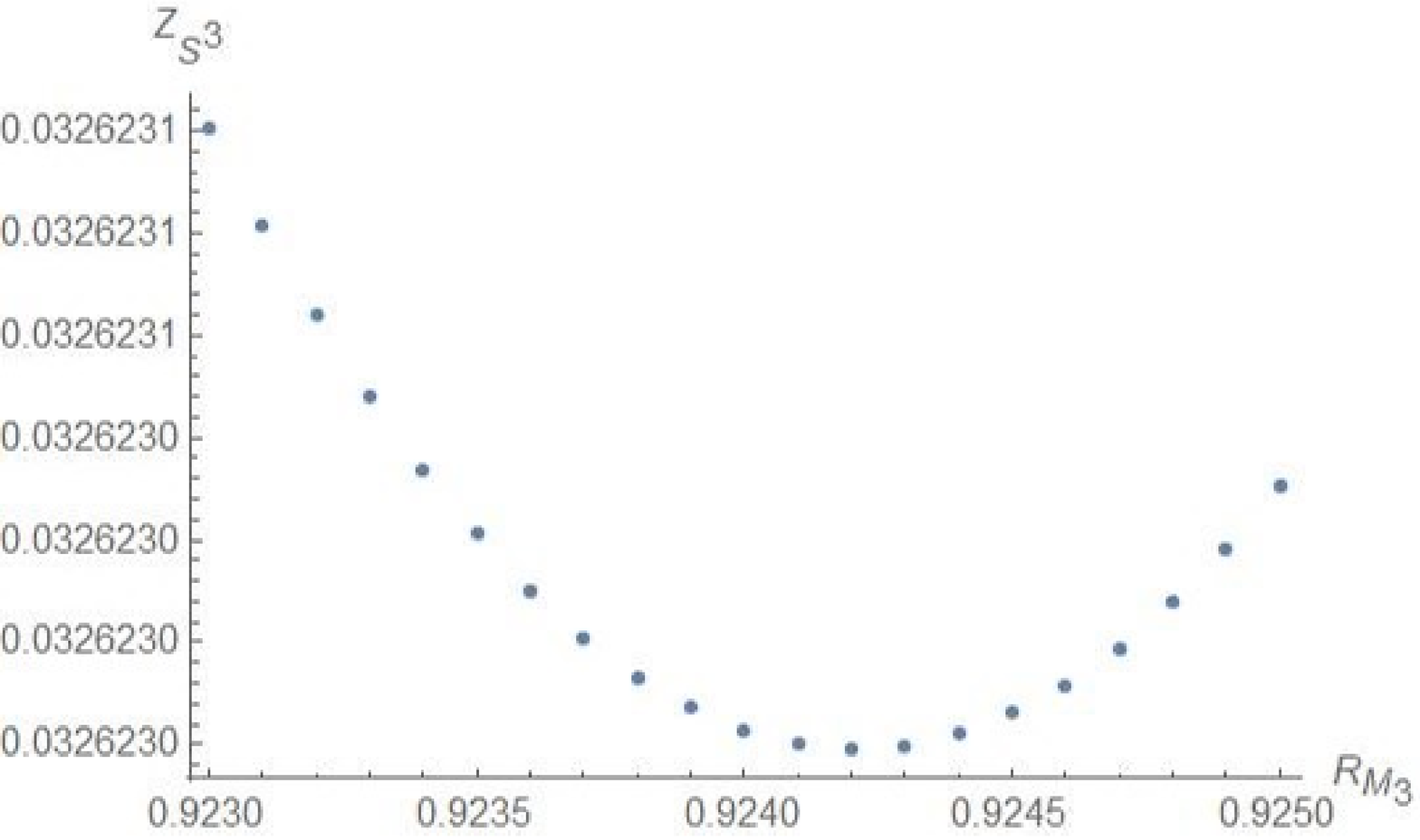}
		\caption{}
	\end{subfigure}
	\begin{subfigure}{5cm}
		\includegraphics[width=5cm]{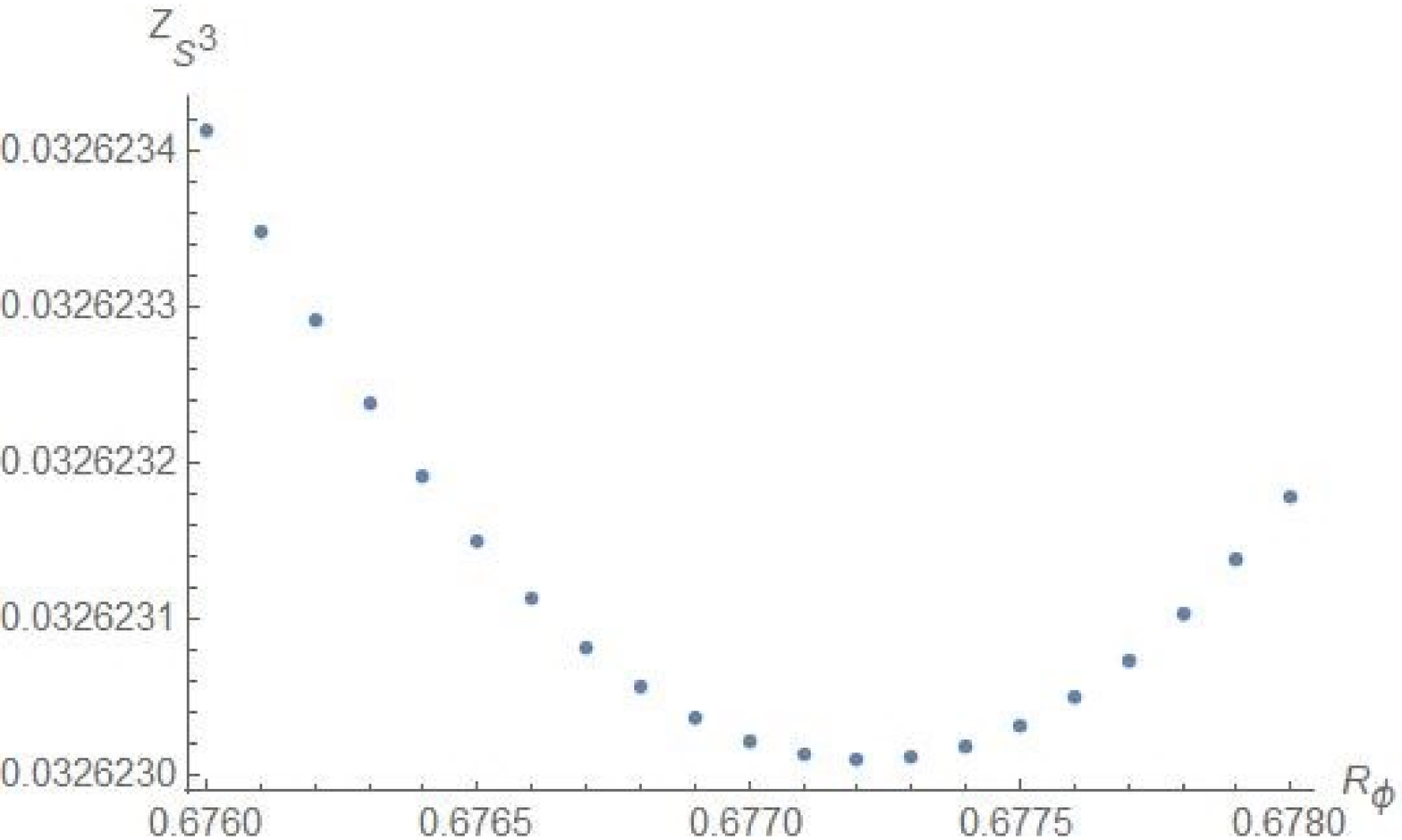}
		\caption{}
	\end{subfigure}	
	\caption{Extremizing the partition function of the $(A_1,D_3)$ Lagrangian including the flipping field $\beta$. The plots give the values after removing the contribution of the decoupled monopole operator. (a) Contour diagram of $\CZ_{S^3}^{(A_1,D_3)+ \beta}$ vs. \ $\{ r_{M_3} , r_{\phi}\}$. (b) Plot of $\CZ_{S^3}^{(A_1,D_3) + \beta}$ vs. \ $r_{M_3}$ at $r_\phi = 0.6772  $. (c) Plot of $\CZ_{S^3}^{(A_1,D_3)+ \beta}$ vs. \ $r_{\phi}$ at $r_{M_3}=0.9242$.   }
	\label{fig:AfterMonDecop}
\end{figure}

\noindent {\textbf{The 3d $(A_1,D_3)$ theory with the flipping field $\beta$}} : A priori, one might expect that if we consider $\mathcal{Z}$-extremization of the theory with the flipping field $\beta $ included in the Lagrangian, then the partition function would extremize at a point where the R-charge for $\phi$ is greater than $\frac{3}{4}$, thereby forcing the R-charge and hence the scaling dimension of $\beta$ to be less than $\half$, hence signaling the decoupling of $\beta$ from the interacting theory. If this were to be the case, it will provide a natural way to explain how $\beta$ gets removed from the 3d Lagrangian. Unfortunately, explicit computations show that this is not the case. Initially the partition function minimizes at $r_{M_3} \simeq 0.9508 $ and $r_\phi \simeq 0.6752$ (see Figure \ref{fig:BeforeMonDecop}). 
At this point in the space of R-charges, the R-charge of the monopole operator is $r_{\mathfrak{M}} = \frac{r_{M_3}}{2} < \half $. Thus the monopole operator decouples as a free chiral multiplet. We will therefore have to factor out the contribution of the decoupled monopole operator and re-extremize the partition function \cite{Morita:2011cs,Agarwal:2012wd,Safdi:2012re}. Upon doing so, we find that the point of extremum is now given by $r_{M_3} \simeq 0.924 $ and $r_\phi \simeq 0.677$ (see Figure \ref{fig:AfterMonDecop}).   
Thus, the  $R$-charge of $\beta$ is given by $r_{\beta} = 0.646$ which is safely above the unitarity bound. Clearly, $\beta$ does not get decoupled from the theory and has to be removed by hand in order to obtain a 3d reduction of the $(A_1,D_3)$ Lagrangian, which is consistent with the 3d reduction of the $(A_1,D_3)$ Argyres-Douglas theory. We also observe that the R-charges at the extremum do not belong to the set of half-integers, thereby ruling out the possibility of SUSY enhancement to 3d $\nn{4}$ at the fixed point, in the Lagrangian where the flipping field is included in the dimensionally reduced theory. 
  
\vspace{5pt}  
  
\subsection{The 3d $(A_1,D_3)$ theory with the flipping field $\beta$ and a monopole superpotential $\beta \mathfrak{M}$}
\label{sec:MonSup}

Note that, in the $(A_1,D_3)$ Lagrangian, the adjoint field $\phi$ contributes 2 fermionic zero modes to the monopole \cite{Nii:2014jsa}. Therefore a monopole superpotential can only be generated if these zero modes can be soaked up appropriately. In the absence of the flipping field $\beta$, there is nothing that can soak up these zero modes and therefore a monopole superpotential will not arise. However, if we do include the flipping field $\beta$ and couple it to the theory through the superpotential given in \eqref{eq:A1D3SuperPot}, then a monopole superpotential of the form 
\be
\delta W = \beta \mathfrak{M}
\ee 
can get generated. 

\begin{figure}[t]
	\centering
	\includegraphics[width=5cm]{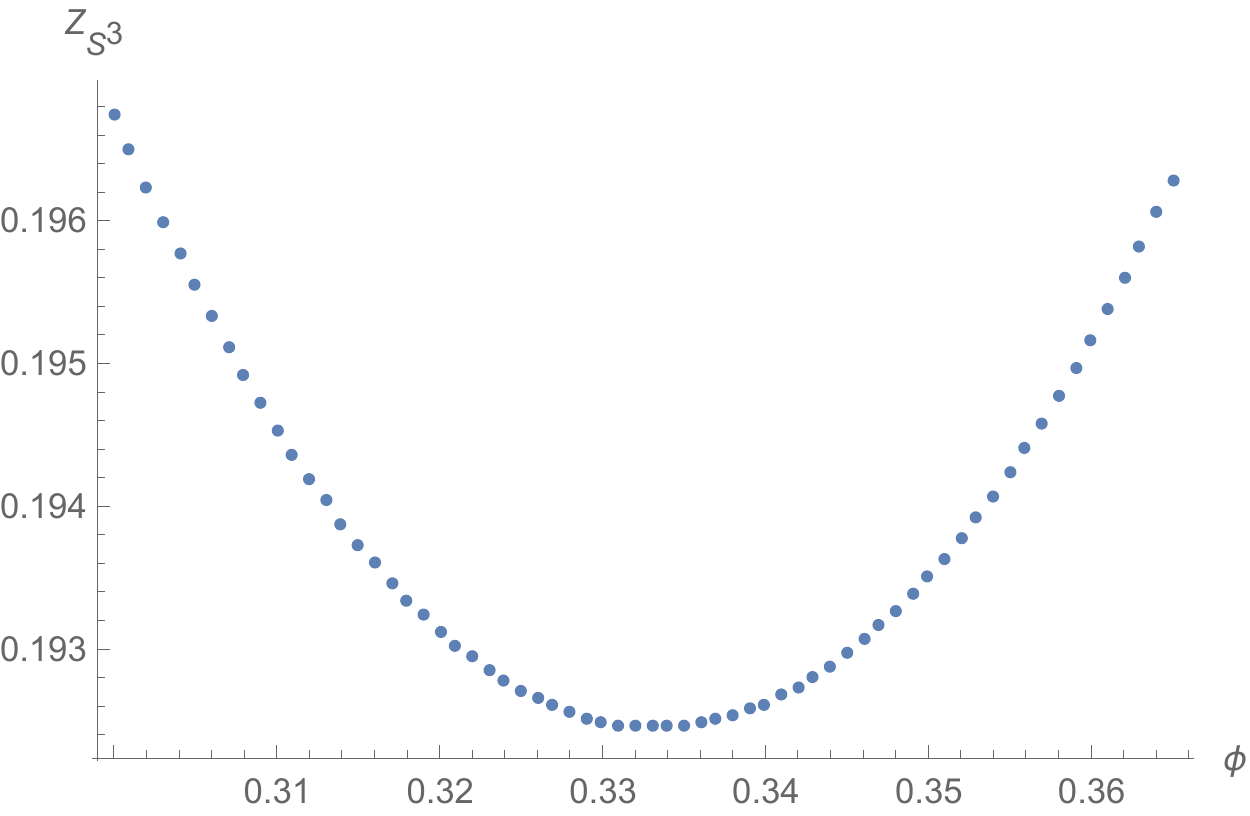}
	\caption{Plot of $\CZ_{S^3}^{(A_1,D_3)+ \beta}$ vs. \ $r_{\phi}$ after including the Monopole superpotential $\delta W =\beta \mathfrak{M}$}
	\label{fig:MonSup}
\end{figure}

We had so far assumed that such a monopole superpotential does not get generated in the 3d reduction of the $(A_1,D_3)$ theory even when the flipping field $\beta$ is included. This was based on the assumption that much like in 4d, the flipping field $\beta$ will not be a part of the 3d chiral ring. Let us loosen this assumption and consider the consequences of assuming that a monopole superpotential does get generated. 
Switching on the monopole superpotential introduces a new constraint on the IR R-charges :
\be
\label{eq:MonSupConstr}
r_{\mathfrak{M}} + r_{\beta} =2 \ . 
\ee    
This forces $r_{M_3} = 4 r_{\phi}$. Thus the the IR R-charges are now parametrized by a single parameter $r_\phi$. We show the plot of $\mathcal{Z}_{S^3}$ vs $r_{\phi}$ in Figure \ref{fig:MonSup}.   
The partition function now extremizes at $r_\phi \simeq \frac{3}{10}$, with all chiral operators being safely above the unitarity bound. One can also check that the R-charges of the gauge invariant operators of the theory do not belong to the set of half-integers. The 3d fixed point therefore is a strictly $\nn{2}$ SCFT. Considering the possibility of a dynamically generated monopole superpotential, therefore does not really bring the 3d Lagrangian to the same fixed point as $T[SU(2)]$.   

\section{Mirror of $(A_1, D_3)$ from the 3d quiver based on the affine $D_4$ Dynkin diagram }
\label{sec:Mirror}

As was shown in \cite{Agarwal:2016pjo}, 4d Lagrangians for the $(A_1,D_{3})$ and $(A_1, D_4)$ AD theories can be obtained from appropriate nilpotent deformations of the 4d $\nn{2}$ $SU(2)$ gauge theory with 8 half-hypers \cite{Maruyoshi:2016aim,Agarwal:2016pjo}. As was argued in \cite{Benvenuti:2017bpg}, we can obtain the 3d mirror of the $(A_1,D_4)$ theory by starting with the mirror of 3d $\nn{4}$ $SU(2)$ gauge theory with 8 half-hypers and then turning on monopole superpotential terms which are the mirror equivalent of the original nilpotent deformations.
 In this section we wish to replicate this procedure for the case of $(A_1,D_3)$ and understand how can we subsequently reduce to the $T[SU(2)]$ theory.  

\subsection{Mirror of 3d $\nn{4}$ $SU(2)$ gauge theory with 8 half-hypers}

\begin{figure}[h]
	\centering
	\includegraphics[width=5cm]{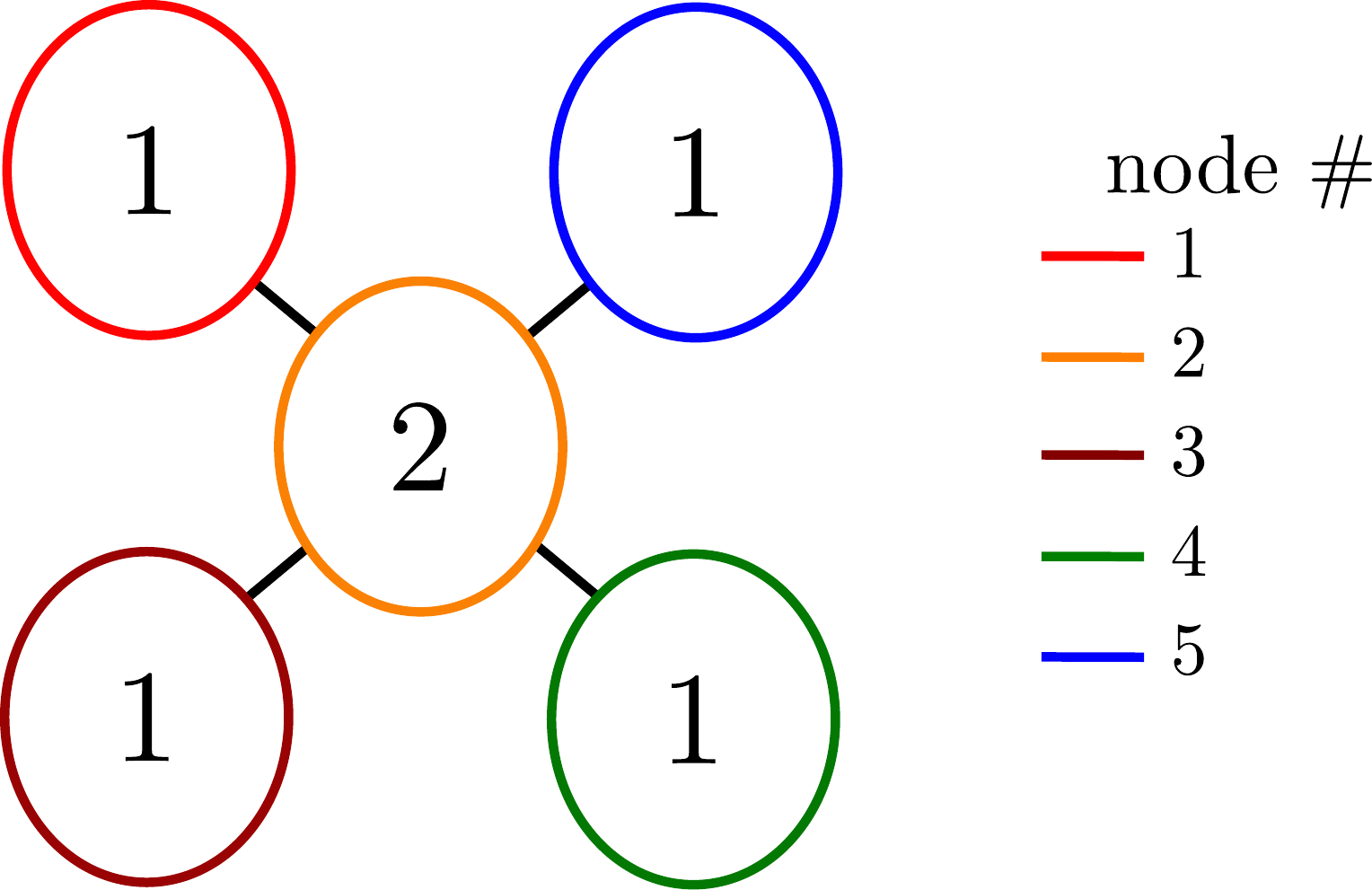}
	\caption{The 3d $\nn{4}$ quiver based on the affine $D_4$ Dynkin Diagram. We number the nodes in the quiver as per the coloring scheme shown in the diagram.}
	\label{fig:D4quiver}
\end{figure}

The mirror for 3d $\nn{4}$ $SU(2)$ gauge theory with 8 half-hypers is given by a 3d quiver, based on the affine $D_4$ Dynkin diagram (see figure \ref{fig:D4quiver}) \cite{Intriligator:1996ex}.
 For the sake of brevity, we will refer to the $SU(2)$ SQCD as the electric theory and the $D_4$ quiver as the magnetic theory. The manifest $SO(8)$ flavor symmetry on the electric side corresponds to the $SO(8)$ generated from the enhancement of the $U(1)^4$ topological symmetry of the magnetic theory. The additional conserved currents required for this enhancement live in the SUSY multiplets of monopole operators with R-charge 1. These were described in section 3.4  of \cite{Cremonesi:2013lqa}. As we will need them later, we will describe these monopole operators here.  The gauge group of the magnetic theory is given by \bigg($U(1)^4 \times U(2) \bigg)/U(1)$. The monopole operators are labeled by a 6-tuple of integers $(m_1,m_2,m_3,m_4,m_5,m_6)$; 
with $m_1, m_4,m_5$ and $m_6$ being the magnetic charges with respect to gauge transformations under $U(1)^4$ and $(m_2,m_3)$ being the magnetic charges with respect to the $U(2)$ gauge symmetry.
The R-charge of generic monopole operator is given by  
\be
\Delta_{\vec{m}} = -|m_2-m_3| + \half \sum_{i \in \{1,4,5,6 \}} \bigg(|m_i-m_2|+ |m_i-m_3|\bigg) \ .
\ee 
From the above formula one can see that monopoles with charge $(n,n,n,n,n,n)$ always have $\Delta=0$. This is precisely because such monopoles are only charged (magnetically) with respect to the over all $U(1)$ which decouples in the magnetic theory. Thus such monopoles are not part of the spectrum of our magnetic theory. It is also implies that $\Delta_{\vec{m}}$ is invariant under a common shift of all the magnetic fluxes $m_i$. One way to fix this freedom is to choose all monopole operators in the theory to have $m_6 = 0$. Furthermore, the action of $U(2)$ Weyl transformations implies that we can always choose to be in the Weyl chamber with $m_2 > m_3$. The topological charges of the monopole operator with flux $\vec{m}|_{m_6=0}$ are then given by $(t_1,t_2,t_3,t_4) = (m_1,m_2+m_3,m_4,m_5)$  . We now look for monopole operators with $\Delta =1$. It can be checked that there are exactly 24 distinct solutions of the above constraints. We explicitly list the monopole operators with positive topological charges in \eqref{tab:D4monopoles}. For each monopole operator listed in \eqref{tab:D4monopoles} with charge $(t_1,t_2,t_3,t_4)$, there is another monopole operator with $\Delta=1$ and topological charge $(-t_1,-t_2,-t_3,-t_4)$ . 
\be
\centering
\begin{tabular}{|c|c|c|c|c|}
	\hline
	mon. op. & $\widetilde{U(1)}_{1}$ & $\widetilde{U(1)}_{2}$ & $\widetilde{U(1)}_{3}$ & $\widetilde{U(1)}_{4}$\\
	\hline \hline
	$\mathfrak{M}_{e_1+e_2}$& 1 & 2 & 1 & 1 \\ \hline
	$\mathfrak{M}_{e_1+e_3}$& 1 & 1 & 1 & 1 \\ \hline
	$\mathfrak{M}_{e_1-e_4}$& 1 & 1 & 1 & 0 \\ \hline
	$\mathfrak{M}_{e_1+e_4}$& 1 & 1 & 0 & 1 \\ \hline
	$\mathfrak{M}_{e_1-e_3}$& 1 & 1 & 0 & 0 \\ \hline
	$\mathfrak{M}_{e_1-e_2}$& 1 & 0 & 0 & 0 \\ \hline
	$\mathfrak{M}_{e_2+e_3}$& 0 & 1 & 1 & 1 \\ \hline
	$\mathfrak{M}_{e_2-e_4}$& 0 & 1 & 1 & 0 \\ \hline
	$\mathfrak{M}_{e_3-e_4}$& 0 & 0 & 1 & 0 \\ \hline
	$\mathfrak{M}_{e_2+e_4}$& 0 & 1 & 0 & 1 \\ \hline
	$\mathfrak{M}_{e_3+e_4}$& 0 & 0 & 0 & 1 \\ \hline
	$\mathfrak{M}_{e_2-e_3}$& 0 & 1 & 0 & 0 \\ \hline
\end{tabular}
\label{tab:D4monopoles}
\ee
Here, $\widetilde{U(1)}_i$ is the topological symmetry associated to the $i$-th node of the $D_4$ quiver as shown in figure \ref{fig:D4quiver}.
As was pointed out in \cite{Cremonesi:2013lqa}, the topological charges listed in \eqref{tab:D4monopoles} coincide exactly with the weight vectors of the $SO(8)$-positive roots in the so called $\alpha$-basis \cite{Feger:2012bs}.  We therefore label the monopoles operators with the corresponding $SO(8)$-roots.

\subsection{Mirror of $(A_1,D_3)$ nilpotent deformation}
\label{sec:MirrorA1D3nil}

The 4d $(A_1,D_3)$ Lagrangian can be obtained by starting with the $SU(2)$ SQCD with 8 half-hypers and considering the $\nn{1}$ nilpotent deformation labeled by $\rho: SU(2)_{\rho} \hookrightarrow SO(8)_{\rm flavor}$ corresponding to the partition $\mathbf{8} \rightarrow \mathbf{5} \oplus \mathbf{1} \oplus \mathbf{1} \oplus \mathbf{1} $ \cite{Agarwal:2016pjo}. This deformation switches on a mass term for some of the half-hypers. Equivalently, it corresponds to deforming the superpotential by 
\be
\delta W = \mu_{j=1, m=-1} \ + \sum_j M_{j,m=-j} \ \mu_{j=j, m=j} \ ,
\label{eq:supdeform}
\ee 
where $ \mu$ is the $SO(8)$ moment map operator and $(j,m)$ are the quantum numbers of the $SO(8)$ adjoint representation with respect to the $SU(2)_\rho \hookrightarrow SO(8)$. For the $SU(2)_\rho$ embedding specified by $\mathbf{8} \rightarrow \mathbf{5} \oplus \mathbf{1} \oplus \mathbf{1} \oplus \mathbf{1} $, the $SO(8)$ irreps. decompose into irreps of $SU(2)_{\rho}$ according to  
\be
SO(8) &\rightarrow& SU(2)_{\rho} \times SO(3)_b \noindent \\ 
\vec{8} &\rightarrow& (\vec{5},\vec{1}) \oplus (\vec{1},\vec{3}) \\
\vec{{\rm adj}} &\rightarrow& (\vec{7},\vec{1}) \oplus (\vec{3},\vec{1}) \oplus (\vec{1},\vec{3}) \oplus (\vec{5},\vec{3})
\ee 
where $SO(3)_b$ is the commutant of $SU(2)_\rho \hookrightarrow SO(8)$. Turns out of all the gauge singlets $M_{j,m=-j}$, only $M_{j=3,m=-3}$ stays coupled to the theory while the others decouple along the RG flow. The singlet field $M_3$ listed in \eqref{tab:A1D3}, is in fact $M_{j=3,m=-3}$ here. In view of the fact that all other gauge singlets decouple, we will remove them from the superpotential deformation of \eqref{eq:supdeform}, which then becomes
\be
\delta W = \mu_{j=1, m=-1} \ + M_{3} \  \mu_{j=3, m=3} \ ,
\label{eq:supdeform2}
\ee
We now consider the 3d mirror version of this deformation. In the mirror theory the deformation \eqref{eq:supdeform2} corresponds to switching on a superpotential term given by the monopole operators dual to $\mu_{j=1, m=-1}$ along with another superpotential term that couples the singlet field $M_3$ to the monopole operators dual to $\mu_{j=3, m=3}$ . These can be easily identified by observing that at the level of $SO(8)$ Lie algebra, $\mu_{j=1, m=-1}$ corresponds to the lowering operator, $X^-$, of $SU(2)_{\rho} \hookrightarrow SO(8)$. The monopole operators dual to $\mu_{j=1, m=-1}$ therefore correspond to the $SO(8)$ roots that together make up the afore mentioned lowering operator. An algorithm to assign an explicit standard triple $\{H, X^+, X^-\}$ for any given nilpotent embedding was given in section 5.2 of \cite{collingwood1993nilpotent}. For the partition given by  $\mathbf{8} \rightarrow \mathbf{5} \oplus \mathbf{1} \oplus \mathbf{1} \oplus \mathbf{1} $, we then obtain 
\be
X^+ &=&  X_{e_2-e_3} + X_{e_3-e_4} + X_{e_3+e_4}, \nonumber \\
X^- &=&  X_{e_3-e_2} + X_{e_4-e_3} + X_{-e_3-e_4}, \nonumber \\
H &=& 4(E_{2,2} - E_{6,6}) + 2(E_{3,3} - E_{7,7}) \ .
\label{eq:stdTriple}
\ee 
Here $X_{\a}$ is the representation matrix for the $SO(8)$-root $\a$ in the fundamental representation.  $E_{i,j}$ is the matrix having 1 as its $(i,j)$-entry and zero elsewhere. Using \eqref{eq:stdTriple} it can also be checked that $e_2+e_3$ is the only $SO(8)$-root that carries $SU(2)_\rho$-quantum numbers $(j=3,m=3)$.   

This implies that the superpotential deformation in the mirror theory is given by 
\be
\delta W = \mathfrak{M}_{e_2-e_3} + \mathfrak{M}_{e_3-e_4} + \mathfrak{M}_{e_3+e_4} + M_3 \ \mathfrak{M}_{-e_2-e_3} \ .
\label{eq:supdeform3}
\ee  
From \eqref{tab:D4monopoles}, we see that $\mathfrak{M}_{e_2-e_3}$ carries a magnetic flux only with respect to the $U(2)$ gauge node of the $D_4$ quiver. Similarly, $\mathfrak{M}_{e_3-e_4}$ only carries a magnetic flux with respect to the $U(1)_3$ gauge node while $\mathfrak{M}_{e_3+e_4}$ only carries a magnetic flux with respect to the $U(1)_4$ gauge node\footnote{The numbering of nodes as described in figure \ref{fig:D4quiver}.}. On the other hand $\mathfrak{M}_{-e_2-e_3}$ carries topological charges $(0,-1,-1,-1)$.

In order to proceed we need to recall that a 3d $\nn{2}$ $U(N_c)$ gauge theory with $N_f = N_c+1$ fundamental flavors and a monopole superpotential $W = \mathfrak{M}^+$ undergoes confinement \cite{Benini:2017dud}. The low energy theory is therefore given by a Wess-Zumino model describing the color singlet particles of the gauge theory. Let $B$ be the $N_f \times N_f$ matrix of $U(N_c)$ singlet particles, and $S$ be a singlet chiral field, then the superpotential of Wess-Zumino model we seek is given by
\be
W = S {\rm det} B \ .
\ee 
We can therefore consider the following 3-step process in our deformed $D_4$ quiver gauge theory
: confinement of $U(1)_3$  $\rightarrow$ confinement of $U(2)$ $\rightarrow$ confinement of $U(1)_4$; we immediately see the $T[SU(2)]$ quiver emerge at the end of the RG flow. 
In the remaining part of this section, we will consider the sequence of afore mentioned steps in more detail and show how they give rise to the superpotential of $T[SU(2)]$. 

After including the superpotential deformation \eqref{eq:supdeform3}, the total superpotential of the quiver gauge theory becomes:
\be
W&=&\sum_{i \in \{1,3,4,5 \}} \phi_i {\rm Tr}(q_i \tilde{q}_i) - \sum_{i \in \{1,3,4,5 \}} {\rm Tr}(q_i \phi_2 \tilde{q}_i) \nonumber \\
&+& \mathfrak{M}_{e_2-e_3} + \mathfrak{M}_{e_3-e_4} + \mathfrak{M}_{e_3+e_4} + M_3 \ \mathfrak{M}_{-e_2-e_3} \ .
\label{eq:D4deformSupPot}
\ee 
Here the fields $(q_i,\tilde{q}_i)$ correspond to the chiral multiplets transforming as the bifundamentals of $U(1)_i \times U(2)$ and $\phi_i$ is the chiral field transforming in the adjoint irrep. of the $i$-th gauge node in the quiver.

\begin{figure}[!htbp]
	\centering
	\includegraphics[width=5cm]{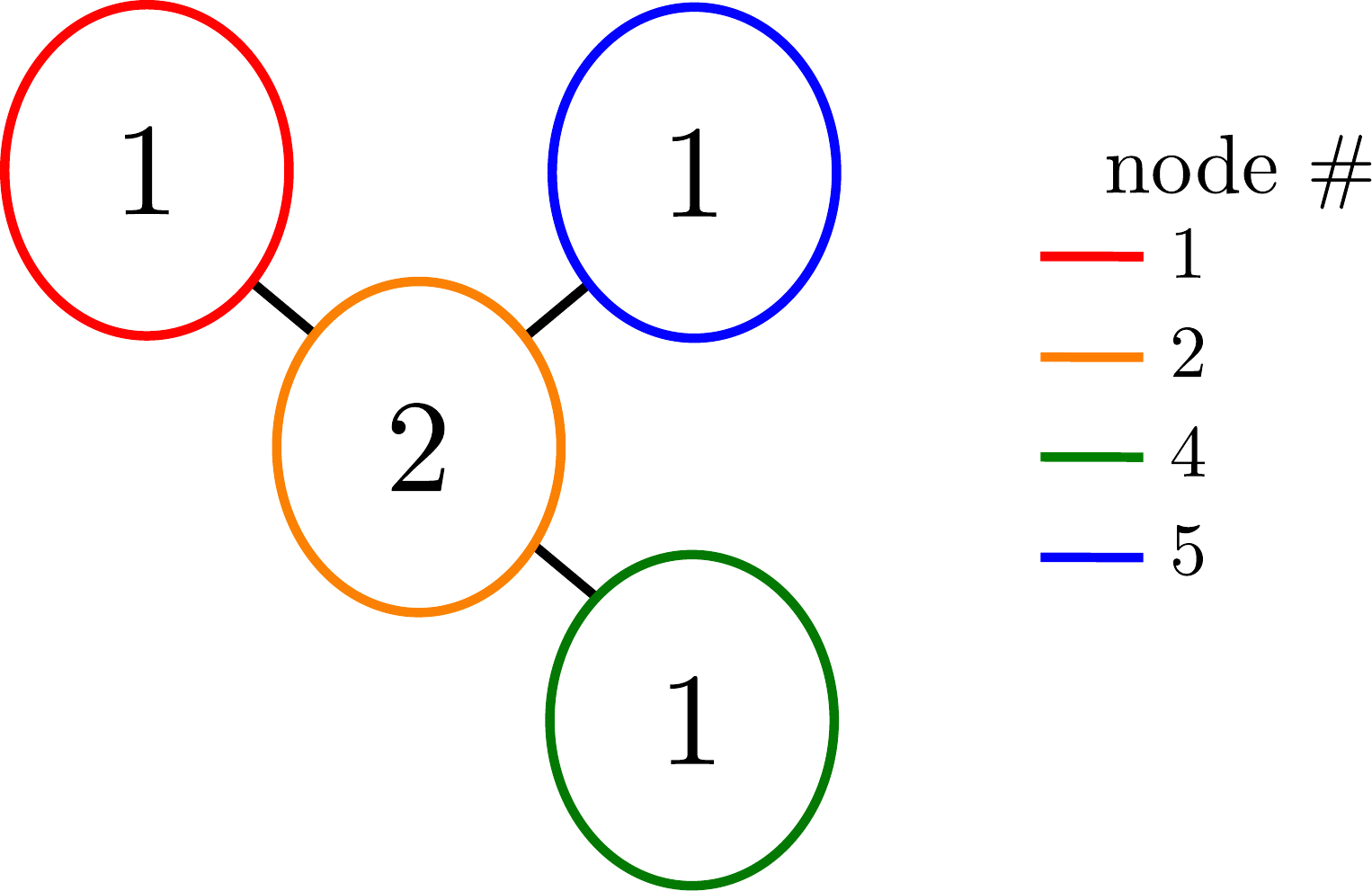}
	\caption{The result of $U(1)_3$ confinement in the $D_4$ quiver of figure \ref{fig:D4quiver}.}
	\label{fig:ConfiningU13}
\end{figure}

\begin{figure}[!htbp]
	\centering
	\includegraphics[width=5cm]{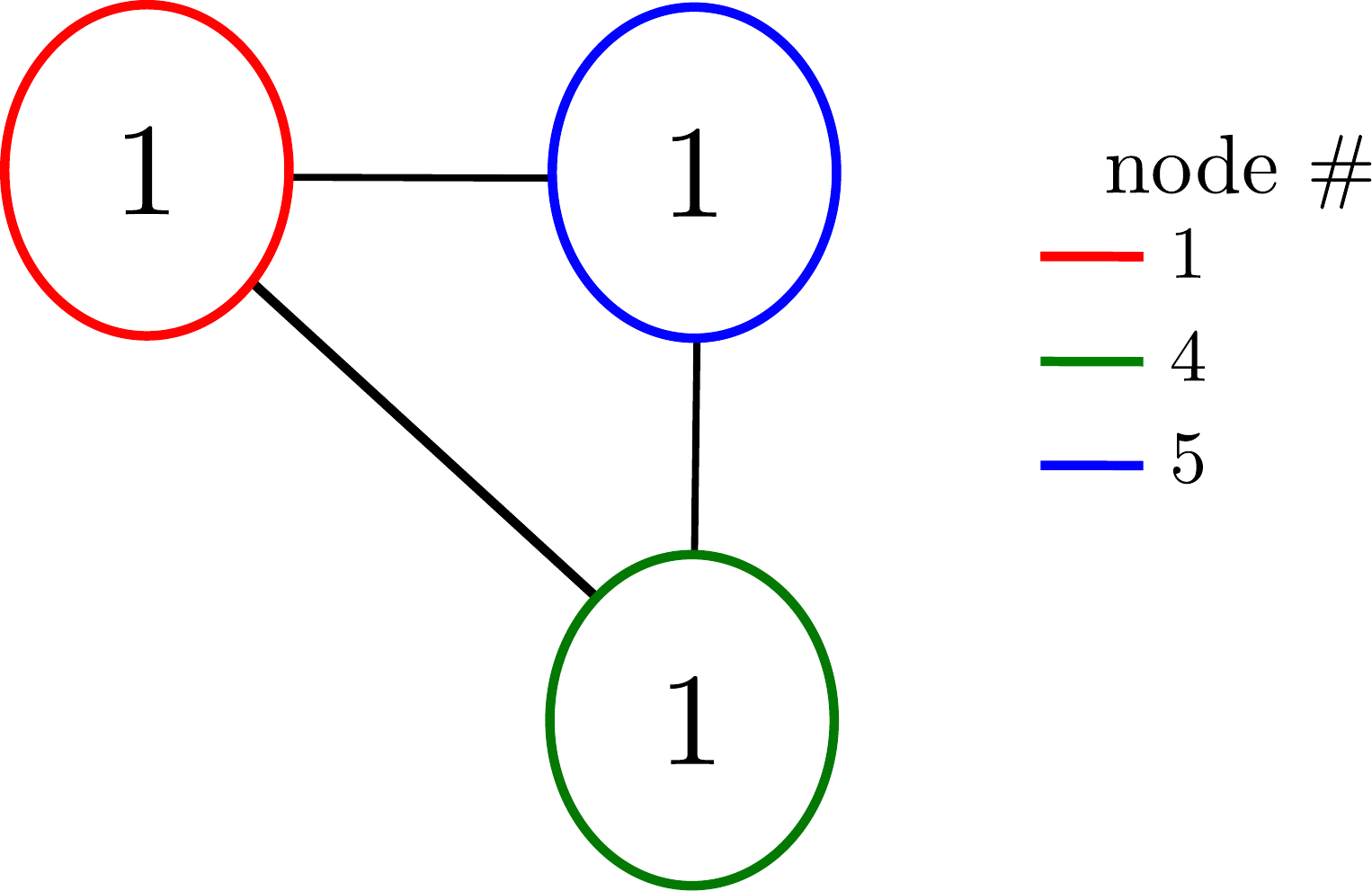}
	\caption{The result of $U(2)$ confinement in the quiver of figure \ref{fig:ConfiningU13}. The bifundamentals of $U(1)_i \times U(1)_j$ gauge nodes are given by $(B_2)_{ij}$ \ . }
	\label{fig:ConfiningU13U2}
\end{figure}

\begin{figure}[!htbp]
	\centering
	\includegraphics[width=5cm]{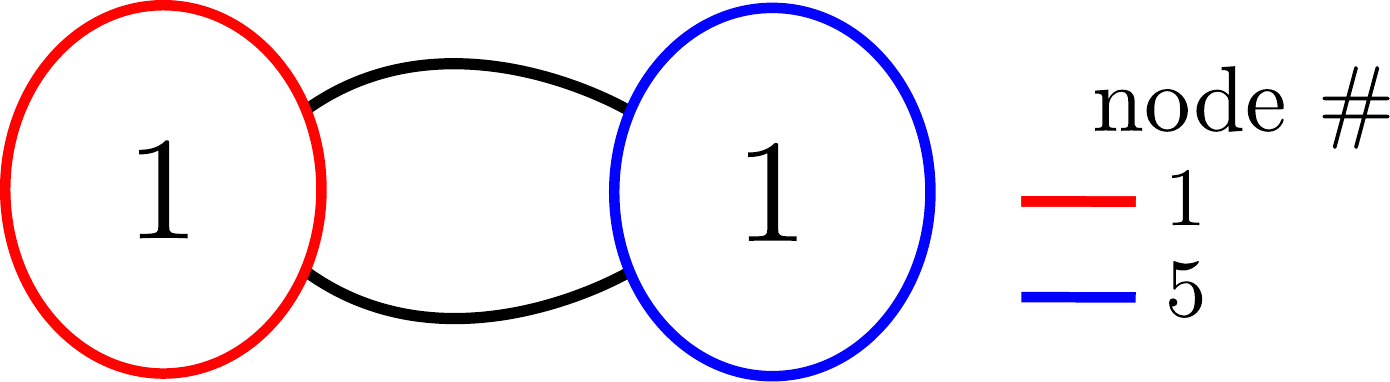}
	\caption{Result of $U(1)_4$ confinement in the quiver of figure \ref{fig:ConfiningU13U2}. The bifundamental hypers are given by $\{(B_2)_{15} , (B_4)_{51}\}$ and $\{(B_2)_{51}, (B_4)_{15}\}$ . }
	\label{fig:TSU2}
\end{figure}

From the local perspective of the $U(1)_3$ gauge node, it is coupled to $N_f=2$ flavors. The presence of the linear monopole superpotential $\mathfrak{M}_{e_3-e_4}$, then implies that the argument of \cite{Benini:2017dud}, is applicable and hence the $U(1)_3$ gauge node confines. This implies that the fields $(q_3,\tilde{q}_3)$ combine to give 4 $U(1)_3$ invariant excitations corresponding the $2 \times 2$ matrix $B_3 = \tilde{q}_3 q_3$. The matrix $B_3$ transforms in the adjoint representation of the  $U(2)$ gauge node. The new quiver resulting from the confinement of $U(1)_3$ is shown in figure \ref{fig:ConfiningU13}.   
The superpotential of the low energy theory now becomes 
\be
W&=&\sum_{i \in \{1,4,5 \}} \phi_i {\rm Tr}(q_i \tilde{q}_i) + \phi_3 {\rm Tr} B_3 -{\rm Tr} (\phi_2 B_3) + \sum_{i \in \{1,4,5 \}} {\rm Tr}(q_i \phi_2 \tilde{q}_i) \nonumber \\
&+&S_3\ {\rm det} B_3  +\mathfrak{M}_{e_2-e_3}  + \mathfrak{M}_{e_3+e_4} + M_3 \ \mathfrak{M}_{-e_2-e_3} \ .
\label{eq:afterU13}
\ee 
The quadratic term ${\rm Tr} (\phi_2 B_3) \subset W$, implies that $\phi_2$ and $B_3$  together become massive and can be intergrated out. The $U(2)$ gauge node is now coupled to exactly $N_f=3$ fundamental flavors. We can therefore apply the arguments of \cite{Benini:2017dud} and conclude that the $U(2)$ gauge node will also confine. The low energy excitations now correspond to $U(2)$ singlets formed from $\{q_i,\tilde{q}_i | i \in {1,4,5} \}$. These can be encapsulated in a $3 \times 3$ matrix $B_2$ such that its $(i,j)$-th entry is given by ${\rm Tr}q_i \tilde{q}_j$ \footnote{We label the rows and columns of $B_2$ by numbers $\{1,4,5\}$. We hope this, slightly unconventional numbering will not cause too much inconvenience to the reader }. The resulting quiver is shown in figure \ref{fig:ConfiningU13U2}.
The superpotential is now given by 
\be
W&=&\sum_{i \in \{1,4,5 \}} \phi_i (B_2)_{i i} + \phi \sum_{i \in \{1,4,5 \}} (B_2)_{ii} \nonumber \\
&+&S_3\ \frac{({\rm Tr} B_2)^2-{\rm Tr}(B_2)^2}{2} + S_2\ {\rm det} B_2   + \mathfrak{M}_{e_3+e_4} + M_3 \ \mathfrak{M}_{-e_2-e_3} \ .
\label{eq:afterU13U2}
\ee  
Where we have used the equation of motion of $\phi_2$ from \eqref{eq:afterU13} to write 
\be
{\rm det} B_3 =  \frac{({\rm Tr} B_2)^2-{\rm Tr}(B_2)^2}{2} \ ,
\ee  
and $\phi:=\phi_3 + {\rm Tr}\phi_2 $ i.e. the linear combination of $\phi_3$ and ${\rm Tr}\phi_2$ that did not get a mass in \eqref{eq:afterU13}. From the form of \eqref{eq:afterU13U2}, it is clear that out the 4 fields $\phi_1,\phi_4,\phi_5$ and $\phi$, exactly 3 linear combinations will get a mass and a fourth linear combination will stay massless. We will use $\hat{\phi}$ to denote the massless linear combination of  $\phi_1,\phi_4,\phi_5$ and $\phi$. Integrating out the massive modes and using their equations of motion we find $(B_2)_{ii}=0, \forall i \in \{1,4,5\}$, the superpotential therefore becomes 
\be
W&=& S_3 \sum_{i<j} (B_2)_{ij}(B_2)_{ji} + S_2 \Big(-(B_2)_{51}(B_2)_{14}(B_2)_{45}+(B_2)_{41}(B_2)_{15}(B_2)_{54}\Big)\ \nonumber \\   
&+& \mathfrak{M}_{e_3+e_4} + M_3 \ \mathfrak{M}_{-e_2-e_3} \ .
\label{eq:afterU13U2simple}
\ee
At this point the $U(1)_4$ gauge node will also confine. The $U(1)_4$ invariant composites formed from $(B_2)_{i4}$ and $(B_2)_{4i}$, give rise to low energy excitations which can be written in the form of a $2 \times 2$ matrix $B_4:(B_4)_{ij} = (B_2)_{i4}(B_2)_{4j}$. The resulting quiver is that of the $T[SU(2)]$ theory as shown in figure \ref{fig:TSU2}.
The low energy superpotential after confinement of $U(1)_4$ becomes 
\be
W&=& S_3 \Big((B_4)_{11}+ (B_4)_{55} +(B_2)_{15}(B_2)_{51}\Big)  + S_2 \Big(-(B_2)_{51}(B_4)_{15}+(B_4)_{51}(B_2)_{15}\Big)\ \nonumber \\   
&+& S_4 ({\rm det} B_4) + M_3 S_4 \ .
\label{eq:afterU13U2U14}
\ee
The fields $S_3, S_4 , (B_4)_{11},(B_4)_{55}$ and $M_3$ are massive and can be integrated out. The superpotential therefore becomes  
\be
W&=&  S_2 \Big(-(B_2)_{51}(B_4)_{15}+(B_4)_{51}(B_2)_{15}\Big)\ ,   
\label{eq:TSU2}
\ee
which matches exactly with the $T[SU(2)]$ superpotential. 
Note that only one linear combination of $(B_4)_{11}$ and $(B_4)_{55}$ gets a mass through superpotential \eqref{eq:afterU13U2U14}, while an orthogonal linear combination decouples as a free field. We identify this with the monopole operator that decoupled from the $(A_1,D_3)$ Lagrangian in section \ref{sec:A1D33dCh}. 
Also note that in arriving at the $T[SU(2)]$ theory starting from the $D_4$ quiver, we did not include the flipping field. This is therefore an independent consistency check of our previous assertion that the flipping field need not be included in the 3d version of the $(A_1,D_3)$ Lagrangian.

\section{Discussion}

In this paper we illustrated an example of a Lagrangian with decoupled operators where the addition of extra-flipping fields seemed to be unnecessary.  Not only this, it appears that upon the addition of a flipping field the expected duality gets violated. However there is a subtle caveat that might be able to explain how even upon the addition of a flipping field, the 3d theory will flow to the $T[SU(2)]$ fixed point in its IR. 
This arises upon considering the possibility of having extra-accidental symmetries on the Coulomb branch of the theory under which the monopole operator $\mathfrak{M}$ acquires non-trivial charges. While the presence of such accidental symmetries is hard to detect right from the onset, if one nonetheless assumes their presence, then these symmetries can mix with the R-symmetry thereby modifying the IR R-charge of monopole operators. If this is indeed the case then, in the scenario where the addition of a flipping field leads to the generation of a monopole superpotential, the constraint $r_{M_3} = 4 r_{\phi}$ resulting from \eqref{eq:MonSupConstr} will be rendered invalid. It can then happen that both, the flipping field $\beta$ as well as  the monopole operator $\mathfrak{M}$ are forced to decouple from the interacting theory in the IR. The IR will then be described by two decoupled sectors: the first one being an interacting non-trivial fixed point while the second being a Wess-Zumino model of two free chiral supermultimets, resulting from the decoupling of $\beta$ and $\mathfrak{M}$, coupled to each other via a mass-term that arises as a result of the monopole superpotential in the UV theory. The sector consisting of the interacting non-trivial fixed point is then expected to be described by the $T[SU(2)]$ theory. Let us also passingly mention that this seems to indicate a certain IR-instability in the theory wherein the presence of the flipping field causes a monopole superpotential to be generated but such a superpotential ultimately causes the flipping field itself to get decoupled from the theory.  

That this might indeed be the case can also be seen if one considers the mirror dual of the 3d theory, as was done in section \ref{sec:MirrorA1D3nil}, with the difference being that this time we also include the flipping field $\beta$ which is now coupled to the rest of the theory via the mirror dual of the electric theory operators ${\rm Tr} \phi^2$ and $\mathfrak{M}$. Both, ${\rm Tr} \phi^2$ and $\mathfrak{M}$ are dual to dimension-2 Higgs branch operators in the mirror theory. As is well known, and is also shown explicitly in appendix \ref{app:HiggsBrD4}, there are only two such independent operators in the $D_4$ quiver. We will denote these by $\mathcal{O}_{13}$ and $\mathcal{O}_{14}$, with $\mathcal{O}_{ij} := {\rm Tr}  \tilde{q}_i q_i \tilde{q}_j q_j  \ ,~\forall  i,j \in \{1,3,4,5\}$. It then follows that in the mirror dual we wish to consider, the superpotential will be given by that presented in \eqref{eq:D4deformSupPot} along with additional superpotential terms given by
\be
\delta W = \beta (a \mathcal{O}_{13} +b \mathcal{O}_{14}) \ ,
\ee
with $a$ and $b$ being some arbitrary constants. The explicit values of $a$ and $b$ will not change the conclusion. What does matter is if $a$ and $b$ are numerically equal to each other.

We can now follow the same steps as in the analysis of section \ref{sec:MirrorA1D3nil}. Upon doing so explicitly, one finds that we once again end up with the quiver shown in figure \ref{fig:TSU2}, however the low-energy superpotential now is given by 
\be
W&=& \beta \Big( (b-a) (B_4)_{11} - a (B_2)_{15}(B_2)_{51}  \Big) + S_3 \Big((B_4)_{11}+ (B_4)_{55} +(B_2)_{15}(B_2)_{51}\Big)\nonumber \\
&+&  S_2 \Big(-(B_2)_{51}(B_4)_{15}+(B_4)_{51}(B_2)_{15}\Big)\ \nonumber \\   
&+& S_4 ({\rm det} B_4) + M_3 S_4 \ .
\ee 
It is easy to see that if the constants $a$ and $b$ are such that $a \neq b$, then the fields $\beta, S_3, (B_4)_{11}$ and $(B_4)_{55}$ are coupled to each other via mass-terms, such that the resulting mass-matrix does not have any null vectors. Hence these get intergrated out. Similarly, $S_4$ and $M_3$ are also coupled by a mass-term and hence get integrated out. The final superpotential therefore once again becomes
\be
W&=&  S_2 \Big(-(B_2)_{51}(B_4)_{15}+(B_4)_{51}(B_2)_{15}\Big)\ ,   
\label{eq:TSU2second}
\ee
thereby, reproducing the $T[SU(2)]$ superpotential.

\acknowledgments
It is a pleasure to thank Antonio Amariti, Dongmin Gang, Seok Kim and Jaewon Song for several helpful discussions. We also wish to thank the anonymous referee whose insightful comments on the previous version of this paper highlighted the role of accidental symmetries in our set-up. These comments were the basis of the discussion section in the current version.  The work of PA is supported in part by Samsung Science and Technology Foundation under Project Number SSTF-BA1402-08, in part by National Research Foundation of Korea grant number 2018R1A2B6004914 and in part by the Korea Research Fellowship Program through the National Research Foundation of Korea funded by the Ministry of Science, ICT and Future Planning, grant number 2016H1D3A1938054.

\appendix
\section{The Higgs branch of the 3d $\nn{4}$ $D_4$ quiver }
\label{app:HiggsBrD4}

The 3d $\nn{4}$ $D_4$ quiver is as given in figure \ref{fig:D4quiver} with the superpotential being 
\be
W&=&\sum_{i \in \{1,3,4,5 \}} \phi_i {\rm Tr}(q_i \tilde{q}_i) - \sum_{i \in \{1,3,4,5 \}} {\rm Tr}(q_i \phi_2 \tilde{q}_i) \ .
\label{eq:D4SupPot}
\ee
For the purpose of our discussion, it will help to keep in mind that $\tilde{q}_i$ are column-vectors while $q_i$ are row-vectors, thus $q_i \tilde{q}_i =  {\rm Tr}(q_i \tilde{q}_i)$ is a color singlet while $\tilde{q}_i q_i$ is a $2 \times 2 $ matrix transforming in the adjoint representation of the $U(2)$ gauge-group associated to the node $\#$ 2 of figure \ref{fig:D4quiver}.   Note that out of the five fields $\phi_1,\phi_3,\phi_4,\phi_5$ and ${\rm Tr}\phi_2$, only 4 independent linear combinations are coupled to the theory while a fifth linear combination does not appear in the superpotential. This is because, an overall $U(1)$ gauge group is decoupled in the quiver with its gauge group being $\Big( U(1)^4 \times U(2) \Big)/U(1)$. The 4 independent linear combinations that stay coupled to the theory can be written as $\varphi_i = \phi_i - {\rm Tr} \phi_2, \ i \in {1,3,4,5}$. The superpotential then becomes 
\be
W&=&\sum_{i \in \{1,3,4,5 \}} \varphi_i {\rm Tr}(q_i \tilde{q}_i) - \sum_{i \in \{1,3,4,5 \}} {\rm Tr}(q_i \varphi_2 \tilde{q}_i) \ ,
\ee 
where $\varphi_2$ denotes the traceless part of $\phi_2$. The F-term equations of motion are now given as
\be
&{\rm e.o.m  \ of \ \varphi_i :}& \  q_i \tilde{q}_i = 0 ~~~~ \forall i \in \{1,3,4,5\} \ , \label{eq:eom1} \\
&{\rm e.o.m  \ of \ \varphi_2 :}& \  \sum_{i \in \{1,3,4,5\}}\tilde{q}_i q_i = 0  \ , \\
&{\rm e.o.m  \ of} \ q_i :& \  \varphi_i \tilde{q}_i -\varphi_2 \tilde{q}_i = 0~~~~ \forall i \in \{1,3,4,5\} \ , \\
&{\rm e.o.m  \ of} \ \tilde{q}_i :& \  q_i \varphi_i -q_i \varphi_2  = 0~~~~ \forall i \in \{1,3,4,5\} \ .
\ee 
Note that the equations of motion of $q_i$ and $\tilde{q}_i$ are automatically satisfied on the Higgs branch since $\vev{\varphi_i} = 0 ~~~ \forall i \in \{1,2,3,4,5\}$ on the Higgs branch. 

Let us now consider the various gauge invariant chiral ring operators on the Higgs branch. We will list them in according to their scaling dimension $\Delta$.

\paragraph{$\Delta=1$}: The only gauge invariant Higgs branch operators with $\Delta=1$ are given by $\mathcal{O}_i=q_i \tilde{q}_i , ~~~ i \in \{1,3,4,5\}$. These are trivial in the chiral ring due to the e.o.m of $\varphi_i$.

\paragraph{$\Delta=2$}: These are given by operators of form $\mathcal{O}_{ij} = {\rm Tr}  \tilde{q}_i q_i \tilde{q}_j q_j  ~~~~ i,j \in \{1,3,4,5\}$. Clearly, $\mathcal{O}_{ij} = \mathcal{O}_{ji}$. It therefore follows that before applying any e.o.m, there are 10 such operators. It is also easy to see that $\mathcal{O}_{ii} = (\mathcal{O}_i)^2 = 0$. Also note that 
\be
\mathcal{O}_{13} + \mathcal{O}_{14} + \mathcal{O}_{15} &=& {\rm Tr} \Big( \tilde{q}_1 q_1  (  \tilde{q}_3 q_3  + \tilde{q}_4 q_4   + \tilde{q}_5 q_5 )\Big) \ , \nonumber \\
                                                          &=& - {\rm Tr} \tilde{q}_1 q_1 \tilde{q}_1 q_1 ~~~~~~~~~~~~ ({\text{by   e.o.m of } \varphi_2 }) \ , \nonumber \\
                                                          &=&  0
\ee   
Similarly, one can establish that $\mathcal{O}_{13} + \mathcal{O}_{34} + \mathcal{O}_{35} = \mathcal{O}_{14} + \mathcal{O}_{34} + \mathcal{O}_{45} = 0$. Additionally, it can be shown that $\mathcal{O}_{13} + \mathcal{O}_{14} + \mathcal{O}_{34} = 0$. One way to see this is to consider the fact that the e.o.m of $\varphi_2$ implies that $ \tilde{q}_1 q_1+ \tilde{q}_3 q_3+  \tilde{q}_4 q_4= -  \tilde{q}_5 q_5$. The above relation then follows from taking the square followed by a trace on both sides of the equality and using the fact that $\mathcal{O}_{ii} = 0  \forall i \in \{1,3,4,5\}$. Using the above relations, we therefore conclude that there are exactly two independent gauge invariant Higgs branch operators with $\Delta=2$ in the chiral ring. Let these be $\mathcal{O}_{13}$ and $\mathcal{O}_{14}$. 

\paragraph{$\Delta=3$}: The gauge invariant Higgs branch operators with $\Delta=3$ are given by 
\be
\label{eq:Op@Del3}
\mathcal{O}_{ijk} = {\rm Tr} \tilde{q}_i q_i \tilde{q}_j q_j \tilde{q}_k q_k   ~~~~ i,j,k \in \{1,3,4,5\}
\ee  
It is easy to show that up to the equation of motion of $\varphi_i$ as given in \eqref{eq:eom1},
\be
\label{eq:Del3const}
\mathcal{O}_{ijk} + \mathcal{O}_{ikj} = 0 \ . 
\ee
The simplest way to do this is to realize that the solution to \eqref{eq:eom1} is given by $q_i^\a \propto \epsilon^{\a \b} \tilde{q}_{i\b}$, substituting this in \eqref{eq:Op@Del3} and using the identity $\epsilon^{\a\b}\epsilon^{\g \r} = \epsilon^{\a\g}\epsilon^{\b\r}-\epsilon^{\a\r}\epsilon^{\b \g}$, we then arrive at \eqref{eq:Del3const} .

Similarly, using the equation of motion of $\varphi_2$, one can show that 
\be
\mathcal{O}_{ijk} + \mathcal{O}_{ijl} = 0, ~~~~ \forall i,j,k,l ~~ {\rm s.t.} ~~ i \neq j \neq k \neq l
\ee 
It therefore follows that the space of chiral operators with scaling dimension $\Delta =3$ has a single generator. Let us choose this to be $\mathcal{O}_{134}$. 

With a little bit of work, it can also be shown that upto equations of motion, 
\be
\label{eq:D4ChiralRel}
(\mathcal{O}_{134})^2 = \mathcal{O}_{13} \mathcal{O}_{14} (\mathcal{O}_{13}+\mathcal{O}_{14}) \ .
\ee 
We therefore arrive at the conclusion that all the chiral operators with scaling dimensions $\Delta \geq 2$ can be written in terms of $\mathcal{O}_{13}$ and $\mathcal{O}_{14}$. Thus the chiral ring of the $D_4$ quiver is generated by two $\Delta=2$ operators : $\mathcal{O}_{13}$ and  $\mathcal{O}_{14}$. 

Upon mirror symmetry, the $D_4$ quiver maps to the $SU(2)$ gauge theory with 4 fundamental hypers, with $\mathcal{O}_{134}$ being mirror dual to the dressed monopole operator $\{\mathfrak{M}\phi\}$. Similarly, the linear combinations $\frac{\mathcal{O}_{13}+\mathcal{O}_{14}}{2^{2/3}}$ and $\frac{\mathcal{O}_{13}-\mathcal{O}_{14}}{2^{2/3}}$ get mapped to ${\rm Tr}\phi^2$ and $\mathfrak{M}$ in the mirror theory. The relation in \eqref{eq:D4ChiralRel} then becomes 
\be
\{\mathfrak{M}\phi\}^2  = ({\rm Tr}\phi^2)^3 - {\rm Tr}\phi^2 \mathfrak{M}^2 \ ,
\ee 
reproducing the Coulomb branch chiral ring relation of 4d $\nn{4}$ $SU(2)$ gauge theory with 4 fundamental hypers. 

\bibliographystyle{jhep}
\bibliography{ADN1}

\end{document}